\newif\ifproceedings\proceedingstrue
\documentclass{IEEEtran}

\newfont{\mycrnotice}{ptmr8t at 7pt}
\newfont{\myconfname}{ptmri8t at 7pt}

\def\maintitle{
SoK: Consensus in the Age of Blockchains}
\def\papertitle{\maintitle} 

\usepackage{subcaption}
\usepackage{times}
\usepackage{hyperref}
\usepackage{breakurl}
\makeatletter
\g@addto@macro{\UrlBreaks}{\UrlOrds}
\makeatother

\usepackage{array,multirow}
\usepackage{xspace}
\usepackage[usenames,dvipsnames]{xcolor}
\usepackage{balance}
\usepackage{paralist}
\usepackage{tabu}
\usepackage{cleveref}
\usepackage[numbers,sort&compress,square]{natbib}
\usepackage{adjustbox}
\usepackage{array}
\usepackage{tabularx}
\usepackage{pifont}% http://ctan.org/pkg/pifont
\usepackage{wasysym}
\usepackage{pdflscape}
\usepackage{adjustbox}
\usepackage{booktabs}
\usepackage{array}
\usepackage{varwidth}
\usepackage{multirow}
\usepackage{tikz}
\usepackage{threeparttable}
\usepackage{rotating}
\usepackage{afterpage}
\usepackage[font=Large,labelfont=bf]{caption}

\usepackage{booktabs}
\newcolumntype{P}[1]{>{\raggedright\arraybackslash}p{#1}}

\definecolor{linkcol}{rgb}{0,0,1}
\definecolor{citecol}{rgb}{0,0.5,0}
\definecolor{urlcol}{rgb}{0.3,0,0}

\renewcommand{\and}{\hspace{5mm}}

\usepackage{enumitem}

\newcommand\comment[1]{}

\newcommand\sarahm[1]{\comment{\textit{\color{OliveGreen}{SarahM: #1}}}}

\newcommand\blockbench{\textsc{Blockbench}\xspace} %bibtex: blockbench

\newcommand\pbft{PBFT\xspace} %bibtex: pbft

\newcommand\bitcoin{Bitcoin\xspace} %bibtex: bitcoin
\newcommand\bitcoinng{Bitcoin-NG\xspace} %bibtex: bitcoinng
\newcommand\ethereum{Ethereum\xspace} 
\newcommand\spectre{Spectre\xspace} 
\newcommand\ghost{GHOST\xspace} 
\newcommand\smartpool{SmartPool\xspace} 
\newcommand\decorehop{DECOR+HOP\xspace} 
\newcommand\fruitchain{Fruitchain\xspace}

\newcommand\permacoin{PermaCoin\xspace} %bibtex: permacoin
\newcommand\spacemint{SpaceMint\xspace} %bibtex: spacemint

\newcommand\cosi{CoSi\xspace} %bibtex: cosi
\newcommand\byzcoin{ByzCoin\xspace} %bibtex: byzcoin
\newcommand\elastico{Elastico\xspace} %bibtex: elastico
\newcommand\omniledger{Omniledger\xspace} %bibtex: omniledger
\newcommand\solidus{Solidus\xspace} %bibtex: solidus
\newcommand\peercensus{Peercensus\xspace} %bibtex: peercensus
\newcommand\algorand{Algorand\xspace} %bibtex: algorand
\newcommand\rscoin{RSCoin\xspace} %bibtex: rscoin
\newcommand\hyperledger{Hyperledger\xspace} %bibtex: hyperledger
\newcommand\chainspace{Chainspace\xspace} %bibtex: chainspace
\newcommand\hybrid{hybrid\xspace} %bibtex: hybrid
\newcommand\randhound{RandHerd\xspace} %bibtex: randhound
\newcommand\bftsmart{BFTSmart\xspace} %bibtex: smartbft
\newcommand\ouroboros{Ouroboros\xspace} %bibtex: ouroboros
\newcommand\ouroborospraos{Ouroboros Praos\xspace} %bibtex: ouroborospraos
\newcommand\snowwhite{Snow-White\xspace} %bibtex: snowwhite

\newcommand{\vs}{vs.\@\xspace}
\newcommand{\etc}{etc.\@\xspace}

\newcommand{\etal}{\textit{et al.}\@\xspace}
\newcommand{\eg}{\textit{e.g.,}\@\xspace}
\newcommand{\ie}{\textit{i.e.,}\@\xspace}
\newcommand{\via}{\textit{via}\@\xspace}

\newcommand\smr{SMR\xspace}
\newcommand\pow{PoW\xspace}
\newcommand\pox{PoX\xspace}
\newcommand\powlong{proof-of-work\xspace}
\newcommand\poxlong{proof-of-X\xspace}

\newcommand\dos{DoS\xspace}
\newcommand\bft{BFT\xspace}
\newcommand\cft{CFT\xspace}

\newcommand\Sybil{sybil\xspace}
\newcommand\sybil{sybil\xspace}
\newcommand\poxStake{proof-of-stake\xspace}
\newcommand\poxElapsed{proof-of-elapsed-time\xspace}
\newcommand\poxCapacity{proof-of-capacity\xspace}
\newcommand\poxRetrieve{proof-of-retrievability\xspace}
\newcommand\poxSpace{proof-of-space\xspace}
\newcommand\vrf{VRF\xspace}
\newcommand\preimage{pre-image\xspace}
\newcommand\stakeholders{stakeholders\xspace}

\def\first{({\it i})\xspace}
\def\second{({\it ii})\xspace}
\def\third{({\it iii})\xspace}
\def\fourth{({\it iv})\xspace}

\newcommand{\marka}{$1$}
\newcommand{\markb}{$2$}
\newcommand{\markc}{$3$}
\newcommand{\markd}{$4$}
\newcommand{\marke}{$5$}
\newcommand{\markf}{$6$}
\newcommand{\markg}{$7$}
\newcommand{\markh}{$8$}
\newcommand{\markj}{$9$}
\newcommand{\markk}{$10$}
\newcommand{\markl}{$11$}
\newcommand{\markm}{$12$}

\newcommand{\cmark}{{\color{OliveGreen}\ding{51}}}%
\newcommand{\xmark}{{\color{BrickRed}\ding{55}}}%
\DeclareRobustCommand\pie[1]{%
    \tikz[every node/.style={inner sep=0,outer sep=0, scale=1.5}]{
        \node[minimum size=1.5ex] at (0,-1.5ex) {}; 
	    \draw[fill=white] (0,-1.5ex) circle (0.75ex); \draw[fill=black] (0.75ex,-1.5ex) arc (0:#1:0.75ex); 
	}%
}

\def\na{--}
\def\unsure{?}
\def\missing{$!$}
\def\L{\pie{0}}	% Low
\def\M{\pie{-180}}	% Medium
\def\H{\pie{360}}	% High

\newcommand\tps{tx/s\xspace}

\begin{document}

\title{\papertitle}

\ifproceedings{
\author{
% \IEEEauthorblockN{ 
% Shehar Bano\IEEEauthorrefmark{1}, 
% Alberto Sonnino\IEEEauthorrefmark{1}, 
% Mustafa Al-Bassam\IEEEauthorrefmark{1},
% Sarah Azouvi\IEEEauthorrefmark{1},
% Patrick McCorry\IEEEauthorrefmark{1},
% Sarah Meiklejohn\IEEEauthorrefmark{1},
% and George Danezis\IEEEauthorrefmark{1}
% }
% \IEEEauthorblockA{\IEEEauthorrefmark{1} University College London, United Kingdom} 

Shehar Bano$^1$,
Alberto Sonnino$^1$,
Mustafa Al-Bassam$^1$,
Sarah Azouvi$^1$,
Patrick McCorry$^1$,\\
Sarah Meiklejohn$^1$,
and George Danezis$^{12}$
\\
\vspace{2mm}
$^1$University College London, United Kingdom\\
$^2$The Alan Turing Institute
}
}\fi

\maketitle

\begin{abstract}

The blockchain initially gained traction in 2008 as the technology underlying \bitcoin~\cite{bitcoin}, but now has been employed in a diverse range of applications and created a global market worth over \$150B as of 2017. 
What distinguishes blockchains from traditional distributed databases is the ability to operate in a decentralized setting without relying on a trusted third party.  
As such their core technical component is \emph{consensus}: how to reach agreement among a group of nodes. 
This has been extensively studied already in the distributed systems community for closed systems, but its application to open blockchains has revitalized the field and led to a plethora of new designs.

The inherent complexity of consensus protocols and their rapid and dramatic evolution makes it hard to contextualize the design landscape.
%It is crucial to develop a systematic understanding of this area, however, because the scalability and performance of blockchains is becoming a bottleneck to their wide-spread adoption. 
We address this challenge by conducting a systematic and comprehensive study of blockchain consensus protocols. After first discussing key themes in classical consensus protocols, we describe: \first protocols based on proof-of-work (\pow), \second proof-of-X (\pox) protocols that replace \pow with more energy-efficient alternatives, and \third hybrid protocols that are compositions or variations of classical consensus protocols. 
We develop a framework to evaluate their performance, security and design properties, and use it to systematize key themes in the protocol categories described above. This evaluation leads us to identify research gaps and challenges for the community to consider in future research endeavours. \sarahm{Should hint at a few if space.}

\end{abstract}

\section{Introduction}

\label{intro}

Blockchains---the technology at the foundation of Bitcoin and other cryptocurrencies---have been hailed as a major disruptive innovation with the potential to transform most industries. The total global market capital of blockchain-based tokens and cryptocurrencies has reached over \$150B as of 2017, and is expected to grow further~\cite{marketcap}. While the projected capabilities and value of the blockchain might seem overly optimistic, its key properties of integrity, resilience, and transparency make it an attractive option for a number of applications. The blockchain is a decentralized, replicated, immutable and tamper-evident log: data on the blockchain cannot be deleted, and anyone can read data from the blockchain and verify its correctness. An important implication of this architecture is \emph{disintermediation}: multiple untrusted or semi-trusted parties can \emph{directly} and \emph{transparently} interact with each other without the presence of a trusted intermediary. This makes blockchains immediately relevant to banks and financial institutions which incur huge middleman costs in settlements and other back office operations. 
A number of big players are actively exploring the feasibility of blockchains, including the Bank of England~\cite{BoE}, the Bank of America~\cite{BoA} and the IMF~\cite{IMF}. However, blockchains are not just restricted to the financial industry; the list of use cases is long~\cite{blockchain-applications}, ranging from voting~\cite{FollowMyVote} and government and public records~\cite{publicRecords1,publicRecords2}, to the sharing economy~\cite{Lazooz,Arcade,LemonWay} and social media~\cite{Akasha,Steem}.    
% \bano{The long list is below}
%other use cases include payments and money transfers~\cite{OpenBazaar}, voting~\cite{FollowMyVote}, intellectual property and digital rights management~\cite{Ascribe,Jaak,Synereo}, sharing economy~\cite{Lazooz,Arcade,LemonWay}, social media~\cite{Akasha,Steem}, supply chain management~\cite{Provenance}, energy management~\cite{TransActiveGrid}, government and public records~\cite{publicRecords1,publicRecords2}, and so on~\cite{blockchain-applications}.

We are at a crucial point in the evolution of blockchains. 
%As stated by Deloitte's Eric Piscini~\cite{blockchain-2017-year}, unless blockchains prove their worth soon, they are bound to hit `enterprise fatigue.' 
The major hurdle in the widespread adoption of blockchains is their performance and scalability---while improvements have been made, they are nowhere near as ubiquitous as their traditional counterparts. These properties are deeply related to the \emph{consensus} protocol---the core component of the blockchain---and we believe this is where future efforts to improve blockchain performance and scalability should be concentrated. The consensus protocol specifies how to get multiple nodes to agree on a value---that is, if a data item should be added to the blockchain. Two key properties of a consensus protocol are: \first requests from correct clients are eventually processed (\emph{liveness}), and \second if an honest node accepts (or rejects) a value then all other honest nodes make the same decision (\emph{safety/consistency}). Consensus is not a new problem: the distributed systems community has extensively studied it for decades, and developed robust and practical protocols that can tolerate faulty and malicious nodes~\cite{pbft,lamport1998part}. However, these protocols were designed for closed groups.

%\begin{figure}[t]
%\includegraphics[width=\linewidth]{graph.pdf}
%\centering
%\caption{\small The number of papers published on %blockchains each year since \bitcoin's inception in 2008 %(sourced from~\cabra~\cite{bibRepo:Decker}).}
%\label{fig:papers}
%\end{figure}

\bitcoin's fundamental innovation was to enable consensus among an open, decentralized group of nodes. This was achieved via a leader election based on proof-of-work (\pow): all nodes attempt to find the solution to a hash puzzle and the node that wins adds the next block to the blockchain. Due to its probabilistic leader election process combined with performance fluctuations in decentralized networks, \bitcoin offers only weak consistency: different nodes might end up having different views of the blockchain leading to \emph{forks}. Additionally, \bitcoin suffers from poor performance which cannot be remedied without fundamental redesign~\cite{croman2016scaling} and its \pow consumes a huge amount of energy~\cite{iceland}. This has led to a plethora of proposals for new consensus protocols~\cite{bano:blockchains:login}.
Some replace \bitcoin's \pow with more energy-efficient alternatives~\cite{permacoin}, while others modify the original design of \bitcoin for better performance~\cite{bitcoinng}.
To achieve strong consistency and similar performance as mainstream payment processing systems like Visa and PayPal, a number of recent proposals seek to repurpose classical consensus protocols for use in decentralized blockchains~\cite{vukolic2016eventually}.
%Each of these proposals modifies or extends the system's safety and liveness guarantees, and under varying assumptions.    

To date there has been no systematic and comprehensive study of blockchain consensus protocols (though there exist a few short surveys based on selected systems which we discuss in Section~\ref{related}). This incurs two major challenges. First, a comprehensive survey of blockchains would doubtless include a discussion of classical consensus protocols. However, the literature is vast and complex, which makes it hard to be tailored to blockchains. Second, conducting a survey of consensus protocols in blockchains has its own difficulties. Though young, the field is characterised by high-volume, fast-paced work. Since 2014, on average about 250 papers per year have appeared on the topic of blockchains. %(Figure~\ref{fig:papers}). 
A reasonable approach is to only consider work published in reputable venues, but here the bulk of the work is published in non peer-reviewed venues and as white papers for industrial platforms. 

We fill this gap by making three contributions. First, we conduct a comprehensive survey mapping how consensus protocols have evolved from the classical distributed systems use case to their application to blockchains. We first discuss key themes in classical consensus protocols (Section~\ref{classical}), and then shift focus to \pow approaches popularized by \bitcoin (Section~\ref{pow}). Section~\ref{pox} discusses proof-of-X (\pox) schemes, which is an umbrella term for systems that replace \pow with more useful and energy-efficient alternatives. In the next two sections, we look at hybrid systems based on novel compositions of classical consensus primitives, or that combine classical consensus with \pow or \pox (Sections~\ref{hybrid-single}~and~\ref{hybrid-multi}). Our second contribution is a common evaluation framework to visualize the capabilities of blockchain consensus protocols (Table~\ref{tb:eval}). Instead of considering individual protocols which would be clearly infeasible, we map out the landscape by extracting and evaluating high-level design themes in blockchain consensus protocols. Finally, we present a discussion of open research challenges and potential directions in the design of future blockchain consensus protocols (Section~\ref{discussion}).

\section{Background and Related Work}

\label{background-related}

%We provide background on blockchains (Section~\ref{background}) and then discuss related work (Section~\ref{related}). 

\subsection{Background}

\label{background}

We describe key concepts in blockchains. For details, we refer the readers to the excellent work by Bonneau~\etal~\cite{bonneau:SOKbitcoin}.  
The \emph{blockchain} is a decentralized, replicated, resilient and transparent data store that allows anyone to read data and verify its correctness. In \emph{permissioned} blockchains, all the node identities are known (trsuted or semi-trusted), and are controlled by a single entity or federation. \emph{Permissionless} blockchains are fully decentralized and anyone can run a node and join the network. 
Data is stored on the blockchain as \emph{blocks}. The blockchain is typically implemented as a linked list in which pointers to previous blocks have been replaced with the cryptographic hash of the previous block. The hash serves as the id of the previous block, and also verifies its integrity. This pattern is repeated in each block, resulting in a \emph{hash chain} in which each block implicitly verifies integrity of the entire chain before it, and tampering with previous data is detectable. It is also possible to store the blockchain as a tree-like structure called the \emph{hash tree} or the \emph{Merkle tree}~\cite{Merkle}. A \emph{transaction} specifies some transformation on the state of the blockchain. If a transaction passes validity and verification checks (\emph{transaction validation}), it is included in a candidate block (a set of transactions) to be added to the blockchain. Nodes in the network participate in a collaborative protocol (\emph{consensus}) to agree on whether the block should be added to the blockchain. In probabilistic consensus protocols like \bitcoin's \pow, nodes might end up having different views of the blockchain (\emph{forks}) because of latency in propagation of transactions, and faulty or malicious nodes. A related concept is that of \emph{double-spending} where a transaction consumes an asset which has already been consumed by a previous transaction. 
Consensus protocols might have a \emph{leader} node that coordinates with other nodes to reach consensus, and for appending a final, committed value to the blockchain. The leader is usually effective for an interval called an \emph{epoch} or a \emph{round}. If the epoch expires (or upon a fault), a new leader is elected. \bitcoin transactions can include reference to well-known deterministic scripts that operate on the transaction inputs and produce some outputs. To make the blockchain a general-purpose platform, scripts are being replaced with \emph{smart contracts}. A smart contract is self-executing code that enforces a digital contract.

\subsection{Related Work}

We believe that our study represents the most comprehensive systematic investigation of consensus protocols in blockchains to date. Below we list work that illuminates different subsets of this space and supports our study.

\label{related}

\subsubsection{Surveys and Systematization} 
Bonneau~\etal~\cite{bonneau:SOKbitcoin} present a comprehensive systematization of \bitcoin and other cryptocurrencies.
Narayanan and Clark~\cite{Narayanan:2017} trace the academic pedigree of \bitcoin technical components.
Zohar~\cite{Zohar2017} provides an overview of scalability and security issues in cryptocurrencies especially \bitcoin, emphasizing the role of incentivization in \pow blockchains to enforce security. Cachin and Vukoli{\'c}~\cite{Cachin:2017} discuss key concepts in classical consensus and describe a selection of permissioned blockchains. Vukoli{\'c}~\cite{vukolic2016eventually} emphasizes the weak consistency of \pow systems. He advocates the shift to classical \bft protocols that offer strong consistency, but notes challenges in their scalability which are discussed in a previous paper by the same author~\cite{vukolic2015quest}. Both these papers focus on permissioned blockchains.  
Yli-Huumo~\etal~\cite{yli2016current} conduct a survey (based on 41 papers) of topic trends in blockchain research and find that 80\% of the papers are on \bitcoin and the remaining 20\% are dominated by security and privacy in blockchains. They highlight concrete evaluation criteria and scalability as neglected areas of research. 

\subsubsection{Evaluation}
\blockbench~\cite{blockbench} is a framework for evaluating the security and performance of private blockchains. Their evaluation reveals that due to design gaps, popular blockchains lag far behind traditional database systems when processing traditional data processing workloads. Their recommendations include systematic benchmarking, improved usability, and revitalizing classical database design principles such as modularity, exploiting hardware primitives, sharding, and support for declarative languages. 
Gervais~\etal~\cite{Gervais:2016} present a quantitative framework to evaluate the security and performance of \pow blockchains. They focus on optimal adversarial strategies for double-spending and selfish mining, while accounting for network propagation, different block sizes, block generation intervals, information propagation mechanism, and the impact of eclipse attacks. 
Croman~\etal~\cite{croman2016scaling} present metrics to evaluate the resource costs and performance of \bitcoin with a focus on scalability. They show that even with reparametrization, \bitcoin can only achieve a maximum throughput of 27 \tps with a latency of 12 seconds.

\section{Systematization Methodology}

\label{method}

%\bano{Establish 3 things here: survey was broad enough to reasonably cover the field (covers selection of papers). the choice of representatives is justified (nothing significant was missed by not including the others). the reviewers are competent and the review process is unbiased and reliable (review/evaluation methodology).}

Capturing a longitudinal \emph{and} representative view of a topic as rich as consensus is challenging. 
%This is compounded by the constant flurry of work that is currently being produced in the area of blockchains. 
We describe our methodology for compiling the literature on which this work is based, and describe the review process and the evaluation framework. We consider consensus in classical systems and blockchains separately because there is a significant difference in the maturity of these two fields. 
%Where relevant, we indicate the domain expertise of the reviewers as `knowledgable' and `expert'.  

\subsection{Classical consensus}

The area of classical consensus is well-established and spans decades. Our goal is to present sufficient background on classical consensus to contextualize its subsequent application to blockchains. %This survey was conducted by one knowledgeable reviewer over a period of 3 months. 
The surveyed literature comprised of well-known seminal works in the area, based on their influence and subsequent citations. It was also supported by Cachin~\etal's comprehensive book on this topic~\cite{Cachin:2011:IRS} and Schneider's classical survey on state machine replication~\cite{Schneider:1990:IFS}.

\subsection{Consensus in Blockchains}

Consensus in blockchains is more involved because it is a high-volume, high-churn evolving area of research. 
%Our key insights in this paper come from this part, including an evaluation framework to systematize work in this area. 
%This task was conducted by six reviewers, where one is a domain expert and the other five are knowledgeable.  

\subsubsection{Compiling and Reviewing Survey Material}
%We first describe how we compiled the literature which forms the basis of this work, and then discuss the review process. 
We used a combination of sources to compile our survey material because the bulk of the work in this area originates in non-academic venues, and the usual metrics such as impact (the number of citations) cannot be employed in this young field. We first compiled a seed-list of literature to survey based on a creditable, actively maintained bibliographic repository on blockchain research by Christian Decker from ETH 
Z{\"u}rich.\footnote{Other similar repositories by Aljosha Judmayer~\cite{bibRepo:Judmayer} and Brett Scott~\cite{bibRepo:Scott} (discontinued after 2016) are also note-worthy; Decker's collection overlaps with both of these.} 
We augmented this list with work that cited peer-reviewed papers in it, and with other relevant papers of which we were aware. 
%Additionally, to keep the list up-to-date we regularly checked Decker's repo, followed blockchain-related profiles and discussions on Twitter, and followed web pages of active blockchain research groups.
%
%In the next step we refined our survey material. Our goal was to extract high-quality papers from the seed-list relevant to consensus in blockchains. (We also extracted surveys on blockchains which have been discussed in Section~\ref{related}.) Our winnowing strategy was: \first discard papers where the title is descriptive and clearly not related to the topic of this paper, \second if the abstract suggests that the paper is relevant, label it by the corresponding category of consensus (\pow, \pox, or \hybrid). 
After further refinement, we ended up with the papers listed in Appendix~\ref{appendix}, which are categorized by PoW, PoX, or \hybrid.
%
%Finally to conduct the review process, the reviewers were split into three groups. Each group was assigned papers in the \pow, \pox, or \hybrid categories. The reviewers held regular meetings to discuss papers in each category with a focus on system design, and the security and performance properties. Next the evaluation framework was developed through iterative discussions. All reviewers had sufficient background knowledge to critically analyze papers in all categories---this enabled inter-rater reliability and helped to verify the soundness of the evaluation process discussed later. 
For each category, a subset of representative papers were selected, prioritizing papers published in academic venues, and papers that significantly advance the field. 

\subsubsection{Evaluation Framework}

Our evaluation framework describes systems along three broad themes: security, performance, and design aspects. 
In terms of security, we consider three properties: \emph{consistency} (\ie whether or not the system will reach consensus on a proposed value), \emph{transaction censorship resistance} (\ie the system's resilience to malicious nodes suppressing transactions), and \emph{\dos resistance} (\ie the system's resilience to DoS attacks against nodes involved in consensus).
In terms of performance, we consider \emph{throughput} (\ie the maximum rate at which values can be agreed upon by the consensus protocol), \emph{scalability} (\ie the system's ability to achieve greater throughput when consensus involves a larger number of nodes) and \emph{latency} (\ie the time it takes from when a value is proposed, until when consensus has been reached on it).
In terms of design, some properties are relevant only to their associated categories, so we defer explanation of them to sections where they are relevant. A complete glossary is included in Appendix~\ref{glossary}.

We use Table~\ref{tb:eval} as a common reference throughout our discussion on \pow, \pox, and \hybrid consensus, focusing on parts of the table relevant to each category.
Unless explicitly stated, we always assume partial synchrony (\ie messages might be delayed in the network but eventually arrive within some bound).
The wide view captured by this table aids in visualizing evaluation of the field. 
%\bano{Note how novel the metrics are. State overlap with other similar work and highlight novel contributions.} 

\section{Classical Consensus}

\label{classical}

%\bano{George: Why are asynchronous consensus protocols like Honeybadger not being used in blockchains? Perhaps mention here, then discuss more in the final Discussion section. See Section 3 in~\cite{vukolic2016eventually} for strong consistency in hardware. ``speeding up strongly consistent services using modern hardware readily available in data centers''.}

%\bano{Say something about optimistic, randomized, and hybrid \bft protocols (see \emph{Eliminating communication and resource overhead in BFT protocols} and \emph{Randomized BFT} in ~\cite{vukolic2015quest})}

%\bano{``the CAP theorem, formally proven in [15], states that in the presence of network partitions, a distributed storage system has to sacrifice either availability or (strong) consistency.''~From~\cite{vukolic2016eventually}}

Safety in distributed systems has been studied since the 1970s, alongside the rise of distributed databases and transactions. Jim Gray, in 1978, proposed the two-phase commit protocol~\cite{gray1978notes}, allowing a transaction manager to atomically commit a transaction, depending on different resources held by a distributed set of resource managers. Transaction commit protocols enable distributed processing, and thus scalability, but do not provide resilience against faulty resource managers, or more generally nodes. In fact, two-phase commit suffers a deadlock in case a resource manager fails to complete the protocol, requiring the introduction of more complex three-round protocols allowing recovery~\cite{skeen1981nonblocking}---i.e.\ the distributed resource managers being able to release the locks held on resources. Since potentially a crucial resource may only be available on a single resource manager, any failures inhibit progress towards accepting dependent transactions.

The need for consensus, or \emph{atomic broadcast}, protocols in distributed systems originates from the need to provide resilience against failures across multiple nodes holding \emph{replicas} of databases. The primitive is closely associated with the \emph{state machine replication} paradigm~\cite{Schneider:1990:IFS} for building reliable distributed computations: any computation is expressed as a state machine, accepting messages to mutate its state. Given that a set of replicas start at the same initial state, and can agree on a common sequence of messages, then they may all privately evolve the state of the computation and correctly maintain consistency across the replicated databases they hold, despite failures or network variations. The underlying consensus protocols are characterized by the communication model, as well as the failure model, assumed.

According to the taxonomy by Dwork~\etal~\cite{dwork1988consensus}, networks may be \emph{syncronous} or \emph{asynchronous}, or offer \emph{eventual synchrony}. In a \emph{synchronous} network the delays messages may suffer can be bound by some time $\Delta$. 
%Thus, strategies based on timers and time-outs (with appropriate exponential back-offs) can be effective in detecting whether a message was sent at all, or not. 
On the other hand, in \emph{asynchronous} networks messages may be delayed arbitrarily, and there exists no reliable bound $\Delta$ for their delay. %Therefore, honest nodes cannot rely on non-reception of messages to infer they were not sent or that they originators are faulty; strategies based on timers and timeouts are ineffective, since they assume that messages should arrive within a certain delay. 
Networks with partially synchronous, or eventually synchronous networks, assume that the network at some stage will eventually be synchronous despite potentially a long period of asynchrony. Fischer~\etal~\cite{fischer1985impossibility} show that deterministic protocols for consensus are impossible in the fully asynchronous case, and have known solutions in the synchronous case (also known as the ``Byzantines General's Problem''). The impossibility theorem is also not taking into account computational bounds on the work nodes may do---something that is exploited by both Nakamoto consensus, as well as other cryptographic solutions~\cite{cachin2000random} to overcome it.

Different failure models have also been considered in the literature. In the \emph{crash failure} model, nodes may fail at any time, but they fail by stopping to process, emit or receive messages. Usually failed nodes remain silent forever, although a number of distributed protocols consider recovery. On the other hand, in the \emph{byzantine failures} model, failed nodes may take arbitrary actions---including sending and receiving sequences of messages that are specially crafted to defeat properties of the consensus protocol. In the network security literature those nodes would be considered malicious or collectively controlled by an adversary. Thus the byzantine setting is of relevance to security-critical settings, and traditional consensus protocols tolerating only crash failures such as Paxos~\cite{lamport1998part}, viewstamped replication~\cite{oki1988viewstamped} and the more modern Raft~\cite{ongaro2014search} or Zab~\cite{junqueira2011zab} cannot be used, unmodified, in adversarial settings.

In terms of the properties expected from a consensus protocol, we consider \emph{liveness} and \emph{safety} as enumerated by Cachin~\etal~\cite{Cachin:2017}. For liveness,  \emph{validity} ensures that if a node broadcasts a message, eventually this message will be ordered within the consensus, and \emph{agreement} ensures that if a message is delivered to one honest node, it will eventually be delivered to all honest nodes. For safety, \emph{integrity} guarantees that only broadcast messages are delivered, and they are delivered only once, and \emph{total order} ensures that all honest nodes extract the same order for all delivered messages.

Consensus refers to `agreement' by all nodes, not 
%`consent' or 
 `choice': consensus protocols are not voting protocols ensuring that all or a majority of nodes agree to the total order, or any single message---the order may be arbitrary or even controlled by an adversary. A number of extensions to consensus protocols include a \emph{validation} step, that ensures the transactions accepted are valid---however the validation rules must be deterministic and uniform across all nodes, and does not afford nodes any discretion about what constitutes a valid message.

An exemplary protocol implementing consensus in the the byzantine and partially synchronous setting is Practical Byzantine Fault Tolerance (PBFT) by Castro and Liskov~\cite{castro01thesis}. The protocol operates in a sequence of views, each coordinated by a leader---a pattern also used in Paxos~\cite{lamport1998part}. Within each view the leader orders messages, and propagates them through a three step reliable broadcast to the replicas. Replicas monitor the leader for safety, as well as for liveness, and can propose a view change in case the leader is unavailable or malicious. Safety is guaranteed within the asynchronous network setting; liveness on the other hand is only guaranteed within a partially synchronous setting, since replicas rely on time-outs to detect a faulty leader. The key complexity of PBFT lies in the view-change sub protocol, that needs to ensure agreement on the new leader and view, as well as guarantee safety of messages agreed in previous views. The basic protocol requires $\mathcal{O}(n^2)$ messages for $n$ replicas to achieve consensus, where $n$ is the number of nodes. The properties of the protocol are guaranteed if $n = 3f+1$, where $f$ is the number of byzantine nodes.

The issue of storage efficiency, a topic of great relevance to blockchain protocols, is discussed in PBFT: a naive implementation of state machine replication, based on consensus, would store the full sequence of actions. The proposed solution relies on replicas agreeing \emph{checkpoint} actions. Those checkpoints are co-signed by all replicas, and allow them to only store the current state of the system, and discard the past sequence that led to the checkpoint state. 

PBFT and other consensus protocols employ replication to achieve resilience against failures, not scalability. In fact the traditional literature on byzantine consensus does not discuss distribution of resources, in the context of a distributed or sharded database, with the exception of a less known joint work by Gray and Lamport on combining atomic broadcast with atomic commit~\cite{gray2006consensus}. As a result, one expects systems employing byzantine consensus to see this protocol become a bottleneck, since its trivial application would require all transactions to be sequenced by the quorum of $n$ nodes---using protocols that are slower than asking a single processor to sequence them.
% ===== Table =====

\newcolumntype{C}{>{\centering\arraybackslash}m{2cm}}
\newcolumntype{B}{>{\centering\arraybackslash}m{1cm}}
\newcolumntype{L}{>{\centering\arraybackslash}V{3cm}}
\renewcommand{\arraystretch}{1.5}

\afterpage{
\clearpage
\begin{landscape}
\centering
\tabcolsep 5pt
\scalebox{.55} {
\begin{threeparttable}
\caption{Evaluation of blockchain consensus protocols. Notation for binary values: \cmark~has property, \xmark~does not have property. Notation for non-binary values: \H~has property, \M~partially has property, \L~does not have property. Notation for meta-information: \na~the property does not apply to the given category,~\unsure~the value could not be extracted,~\missing~the value is missing. The rows correspond to selected systems in each protocol category; a full list of the corresponding citations is provided in Appendix~\ref{appendix}. A list of terms is included in Appendix~\ref{glossary}. In the \emph{Msg.} column (message complexity), $n$ refers to the number of participants, and $c$ is the size of the committee.}
\label{tb:eval}

\begin{tabular}{ c | LB |  *{2}{C} | *{4}{C} | *{3}{C} | *{3}{C} | *{4}{C} }
\toprule
% =========== headers I =========== %
\multicolumn{2}{c}{\bf Systems}
& \bf code avail.
& \bf Committee Formation (Resources) 
& \bf Strong consistency
& \multicolumn{4}{c|}{ \bf
    \begin{tabular}{ *{4}C }
        \multicolumn{4}{c}{Single Committee} \\
        \midrule
        %\multirow{2}{*}{Committee Configuration} & 
        Committee Configuration &\multicolumn{3}{c}{Inter-Committee Consensus} \\
        \cmidrule{2-4} \\
        & Incentives (Join,Participate) & Leader & Msg.
    \end{tabular}
} 
& \multicolumn{3}{c|}{ \bf
    \begin{tabular}{ *{3}C }
        \multicolumn{3}{c}{Multiple Committee} \\
        \midrule
        %\multirow{2}{*}{Node-to-Committee Configuration} &
        Intra-Committee Configuration & \multicolumn{2}{c}{Intra-committee Consensus} \\
        \cmidrule{2-3}
        & Mediated & Incentives
    \end{tabular}
} 
& \multicolumn{3}{c|}{ \bf
    \begin{tabular}{ *{3}C }
        \multicolumn{3}{c}{Safety} \\
        \midrule
        Transaction Censorship Resistance & DoS Resistance & Adversary Model
    \end{tabular}
}
& \multicolumn{4}{c}{ \bf
    \begin{tabular}{ *{4}C }
        \multicolumn{4}{c}{Performances} \\
        \midrule
        Throughput & Scalable & Latency & Exp. Setup
    \end{tabular}
}
\\

\toprule
% =========== hybrid =========== %
\parbox[c]{3mm}{\multirow{8}{*}{\rotatebox[origin=c]{90}{\bf\normalsize hybrid}}}
& \bf \byzcoin~\cite{byzcoin} & \cmark & PoW & \cmark & Rolling (single) & \cmark \xmark & Internal & $O(n)$ & - & - & - & \cmark & \M & 33$\%$  & 1000 \tps \tnote{\marka} & \xmark & 10--20s \tnote{\marka} & Real  \\

& \bf \solidus~\cite{solidus} & \xmark & PoW & \cmark & Rolling (single) & \cmark \cmark & External & $O(n^2)$ & - & - & - & \xmark &  \M & 33$\%$ & - & - & - & - \\

& \bf \algorand~\cite{algorand}  & \xmark & Lottery & \cmark & Full swap & \xmark \xmark & Internal & $O(n^2)$ & - & - & - &  \xmark &  \H & 33$\%$ & 90 tx/h \tnote{\markb} & \xmark & 40s \tnote{\markb} & Real  \\

& \bf \hyperledger~\cite{Vukolic:2017:RPB}  & \cmark & Permissioned & \cmark & Static & - & Flexible & Flexible & - & - & -  & \cmark & \H & 33$\%$ & 110k \tps \tnote{\markc} & \xmark & $<$1s \tnote{\markc} & Real \\

& \bf \rscoin~\cite{rscoin}  & \cmark & Permissioned & \cmark & Static & - & Internal & $O(n)$ & \xmark & Client & \xmark & \cmark & \H & 33$\%$  & 2k \tps \tnote{\markd} & \cmark & $<$1s \tnote{\markd} & Real \\

& \bf \elastico~\cite{elastico}  &  \xmark & PoW & \cmark & Full swap & \cmark \xmark & Internal & $O(n^2)$ & Dynamic (Random) & \missing & \missing & \xmark & \H & 33$\%$ & 16 blocks in 110s \tnote{\marke}
& \cmark & 110s for 16 blocks \tnote{\marke} & Real  \\

& \bf \omniledger~\cite{omniledger}  & \xmark & \pow/\pox & \cmark & Rolling (subset) & \cmark \xmark & Internal & $O(n)$ & Dynamic (Random) & Client & \xmark & \cmark & \H & 33$\%$ & $\approx$10k \tps \tnote{\markf} & \cmark & $\approx$1s \tnote{\markf} & Real \\

& \bf \chainspace~\cite{chainspace}  & \cmark & Flexible & \cmark & Flexible & \xmark \xmark & Internal & $O(n^2)$ & \xmark & \xmark & \xmark  & \cmark & \M & 33$\%$ & 350 \tps \tnote{\markg} & \cmark & $<$1s \tnote{\markg} & Real \\

\midrule
% =========== pox =========== %
\parbox[c]{3mm}{\multirow{7}{*}{\rotatebox[origin=c]{90}{\bf\normalsize\poxlong}}}
& \bf Ouroboros \cite{ouroboros}  & \xmark & Lottery & \xmark & Full swap &\cmark\cmark & Internal & $O(nc)$ &\na &\na &\na&\xmark & \M & 50$\%$  & 257.6 \tps\tnote{\markj} &\xmark & 20s & Simulation\\

& \bf Praos \cite{ouroborospraos}  & \xmark & Stake & \xmark & Rolling (subset) &\cmark\cmark & Internal & O(1) & \na & \na&\na &\xmark & \H & 50$\%$  &\na &\na &\na & \na \\

& \bf Snow-white \cite{snow-white}  & \xmark & Stake & \xmark & Full swap &\cmark\cmark & Internal & O(1) & \na& \na &\na &\xmark & \M & 50$\%$ & 100-150 \tps\tnote{\markj} &\cmark & \unsure & Simulation \\

& \bf PermaCoin \cite{permacoin}  & \cmark & \pow/PoR\tnote{\markl} & \xmark & Rolling (single) & \xmark\cmark & Internal & $O(1)$ & \na & \na & \na & \cmark & \H & 50$\%$ & \na & \xmark & \na & \na \\

& \bf SpaceMint \cite{Fuchsbauer2015SpaceMintAC}  & \cmark & PoS & \xmark & Rolling (single) & \xmark\cmark & Internal & $O(1)$ & \na & \na & \na & \cmark & \H & 50$\%$ & \unsure & \xmark & 600s & Simulation \\

& \bf Intel PoET \cite{sawtooth}  & \cmark & TH\tnote{\markm} & \xmark & Rolling (single) & \xmark\cmark & Internal & $O(1)$ & \na & \na & \na & \cmark & \H & TH\tnote{\markm} & 1000 \tps\tnote{\markk} & \cmark & \na & Real \\

& \bf REM \cite{rem}  & \xmark & TH\tnote{\markm} & \xmark & Rolling (single) & \xmark\cmark & Internal & $O(1)$ & \na & \na & \na & \cmark & \H & TH\tnote{\markm} & ! & \cmark & \na & Real\\

\midrule
% =========== pow =========== %
\parbox[c]{3mm}{\multirow{5}{*}{\rotatebox[origin=c]{90}{\bf\normalsize\powlong}}}
& \bf Bitcoin \cite{bitcoin} & \cmark & \pow & \xmark & Rolling (single) & \xmark\cmark & Internal & $O(1)$ & \na & \na & \na  & \cmark & \H & 50$\%$ & 7 \tps & \xmark & 600s & Real\\

& \bf Bitcoin-NG \cite{bitcoinng} & \xmark & \pow & \xmark & Rolling (single) & \xmark\cmark & Internal & $O(1)$ & \na & \na & \na & \cmark & \M & 50$\%$ & 7 \tps & \xmark & $<$1s & Simulation \\

& \bf GHOST \cite{sompolinsky2013accelerating} & \xmark  & \pow & \xmark & Rolling (single) & \xmark\cmark & Internal & $O(1)$ & \na & \na & \na  &  \cmark & \H & 50$\%$ & \na & \xmark & \na & \na\\

& \bf DECOR+HOP \cite{lernerdecor+} & \xmark  & \pow & \xmark & Rolling (single) & \xmark\cmark & Internal & $O(1)$ & \na & \na & \na & \cmark & \H & 50$\%$ & 30 \tps\tnote{\markh} & \xmark & 60s & Simulation\\

& \bf Spectre \cite{sompolinsky2016spectre} & \xmark & \pow & \xmark & Rolling (single) & \xmark\cmark & Internal & $O(1)$ & \na & \na & \na & \cmark & \H & 50$\%$ & \na & \xmark & \na & \na \\

\bottomrule
\end{tabular}

% =========== footnote =========== %
\begin{tablenotes}
\large
\item [\marka] 144 nodes/committee.
\item [\markb] 50k nodes/committee.
\item [\markc] 4 nodes/committee (corresponding to \bftsmart~\cite{hyperledger-bftsmart}).
\item [\markd] 3 nodes/committee. 10 committees.
\item [\marke] 100 nodes/committee. 16 committees.
\item [\markf] 72 nodes/committee (12.5\% adversary). 25 committees.
\item [\markg] 4 nodes/committee. 15 committees.
\item [\markh] 1 minute average interval; 1 block $=$ 1 MB. 
%\item [\marki] c is the size of the committee.
\item [\markj] 40 nodes.
\item [\markk] As reported in a blog post~\cite{altoros}.
\item [\markl] \poxRetrieve.
\item [\markm] Trusted Hardware.
\end{tablenotes}
\end{threeparttable}
}
\end{landscape}
}

\section{Proof-of-Work Consensus}\label{pow}

In 2008, \bitcoin~\cite{bitcoin} was published by a pseudonymous author Satoshi Nakamoto; it has since gone on to become one of the most successful cryptocurrencies of modern times. The key innovation of \bitcoin is its use of proof-of-work (\pow) to achieve consensus---also called Nakamoto consensus after its originator---in a fully decentralized, permissionless network. %\alberto{To date, \pow consensus are still extremely important as they account for more than 90$\%$ of the market share of digital currencies~\cite{liu2016scalable}.}

% ======

\subsection{Nakamoto consensus}\label{sec:nakamoto_pow}

While the technical components of \bitcoin originate in previous academic literature, their composition in \bitcoin to achieve consensus is novel. The idea of proof-of-work was first presented by Dwork and Naor in 1993 as a technique for combatting spam mail, by requiring the email sender to compute the solution to a mathematical puzzle to prove that some computational work was performed~\cite{dwork1993crypto}.

\pow was independently proposed in 1997 for Hashcash by Back, another system for fighting spam~\cite{hashcash}. In Hashcash, the computational puzzle is finding a SHA-1 hash of a header including the email recipient's address and current date, such that the hash contains at least 20 bits of leading zeros. As the hashing algorithm is \preimage resistant, the puzzle can be solved only by including random nonces in the header until the resulting hash meets the leading zeros requirement. These guesses require a significant amount of computational work, so a valid hash is considered to be a \pow.

Nakamoto consensus is derived from Hashcash~\cite{hashcash}. It replaces Hashcash's SHA-1 hashing with two successive SHA-2 hashes, and requires valid hashes to have a value below a target integer value $t$. The difficulty of the puzzle is therefore adjustable: decreasing $t$ increases the number of guesses (and thus work) required to generate a valid hash. The nodes that generate hashes are called \emph{miners} and the process is referred to as \emph{mining}. Miners calculate hashes of candidate blocks of transactions to be added to the blockchain, and are rewarded with new coins if they find a valid block.
The value $t$ is reset by the network every 2016 blocks such that miners are successful (and can append a block to the blockchain) probabilistically every 10 minutes (also called the \emph{inter-block interval}).

% ======

\subsection{Forks}

Forks in the blockchain may occur if two miners find two different blocks that build on the same previous block. This is resolved by \pow consensus, which orders transactions and makes double-spending expensive. In the original paper, forks are resolved in the consensus rules by accepting the `longest chain, which has the greatest proof-of-work effort invested in it' as the correct one. In practice, this is implemented as the chain with most accumulated work, as it is possible for a shorter chain to have more proof-of-work than a longer chain.

%\subsubsection{The 51\% Attack}

To double-spend assets on a \pow blockchain, an attacker must have sufficient computing power to be able to create a fork of the blockchain that has more accumulated work than the chain that is to be overridden. Thus the threat model assumes an adversary that has the majority of the computing power on the network (referred to as a \emph{51\% attack}) can outpower the remaining computational power in generating a chain with the most accumulated work. The \textit{security threshold} of the network is the percentage of computing power required to conduct a 51\% attack. Decker and Wattenhofer showed that due to the delays in blocks propagation in the \bitcoin network, increasing the block size and decreasing the inter-block interval increases the chance of forks occurring~\cite{conf/p2p/DeckerW13}, as delayed miners may waste effort in attempting to mine on top of blocks that are no longer the latest ones.
%This is because if block propagation is slower for miners who suffer from delays than the time it takes for miners to generate blocks, then miners will attempt to mine on top of blocks that may already have other blocks built on top of them (but failed to reach the miner due to delays). 
%This causes miners to wastefully mine blocks that will be discarded, as only one fork will survive. 
As a result, the network becomes more susceptible to 51\% attacks from a miner that does not suffer from delays.

% =======

\subsection{Scaling \bitcoin}

Bitcoin currently has a hardcoded blocksize limit of 1MB per block, and a 10 minute block frequency target. Gervais~\etal~\cite{Gervais:2016} showed that \bitcoin's block frequency can be reduced to 1 minute per block without reducing the security threshold of the existing network, modelling the bandwidth distribution of the network around real-world broadband data. 

Forks might still occur in \pow blockchains despite countermeasures to avoid them. New policies have been proposed for the selection of the main chain in the forked blockchain to obtain a more resilient and scalable system than \bitcoin.
\ghost~\cite{sompolinsky2013accelerating}
 exploits blocks that are not on the main chain, achieving higher transaction rates without undermining \bitcoin security.
 %particularly, they are able to increase the block generation rate without increasing the susceptibility of a 50$\%$ attacks.
Unlike \bitcoin's linear blockchain, \ghost organizes blocks in a tree structure. The tree is shaped by the blocks that successful miners choose to extend. The chain selection algorithm chooses the heaviest path as main chain, where a block's weight depends on how dense its subtree is.

The challenge of scaling \pow blockchains is that the more one increases throughput (block size) or the more one decreases latency (block frequency), the lower the resilience of the network to 51\% attacks. A common theme in enhancing \pow consensus is to improve performance while maintaining the security threshold of the network without requiring nodes and miners to upgrade their network connections.

\bitcoinng~\cite{bitcoinng} shares \bitcoin's trust model, but improves performance by separating leader election from transaction serialization (\ie appending them to the blockchain). In each epoch, a leader is selected via \pow as in \bitcoin. Unlike \bitcoin, the leader can continue to append transactions to the blockchain for the duration of its epoch, until a new leader is elected. This allows latency to be limited only by the network's propagation delay, and bandwidth to be limited only by the processing capacity of the nodes. Another approach for improving performance, used by \spectre~\cite{sompolinsky2016spectre}, is to allows miners to mine blocks concurrently by replacing the `linear' blockchain structure with a block-DAG. %To avoid double-spending, \spectre specifies a pairwise ordering on blocks such that blocks with the higher ordering are selected. 
Off-chain approaches to improve \bitcoin scalability such as the Lightning Network~\cite{poon2015bitcoin} have also been proposed, where parties can execute transactions off the main consensus path, and submit only the final state to the blockchain. A more detailed discussion of off-chain solutions is outside the scope of this work.

% ======

\subsection{Mining Centralization}

\label{pow:miningCentralization}

To reduce the variance of miners' rewards, miners often aggregate resources and share rewards among themselves \via pooled mining protocols. However, mining pools undermine decentralization and are vulnerable to transaction censorship by a malicious pool manager (needed to map transactions to blocks)~\cite{luu2017smart}. To mitigate such attacks, the \pow mechanism should be \emph{fair}: the number of valid blocks mined by a miner should be proportional to its computing power in the network. A number of techniques have been proposed to create decentralized mining pools~\cite{luu2017smart, miller2015nonoutsourceable}.
\smartpool~\cite{luu2017smart} implements a practical decentralized mining pool through an \ethereum smart contract, with the smart contract replacing the traditional pool manager.  On the other hand, Miller~\etal~\cite{miller2015nonoutsourceable} discourage mining pools by proposing non-outsourceable proof-of-work puzzles, in which rewards can be entirely stolen from the pool manager by the entity solving the puzzle, without producing any evidence of its implication.
%Specifically, they craft the new puzzle such that if a worker is doing a large part of the mining computation, it must possess a sufficiently large part of a “signing key” such that it can later sign over the reward to its own public key

%Another topic is to improve fairness between miners. Such efforts aim to avoid centralization by giving miners the same reward and \bano{Mustafa: The following is not clear} guaranteeing low variance as if they were mining with centralized pools. \decorehop~\cite{lernerdecor+} enforces fairness between miners by allowing miners to share the profit when competing blocks are generated.
\decorehop~\cite{lernerdecor+} enforces fairness between miners by allowing them to share the profit when competing blocks are generated. %\paddy{It shares reward in forks; and punishes miners that do not follow the network's selection rules on blocks (i.e. if i do not extend the heaviest branch; I get punished for that - so the block reward is split between us; and i lose a percentage of my split. If my uncle blocks get into the blockchain late that is also punished.) the insight is that DECOR is a punishment protocol; whereas Ethereum's protocol is a subsidy protocol and Bitcoin just rewards best-effort. } 
Such efforts aim to avoid centralization by giving miners the same reward and guaranteeing low variance as if they were mining with centralized pools. Moreover, \decorehop improves \bitcoin performance by using `header-first propagation', where the block header is sent first, and nodes attempt to reconstruct the full block from transactions that they have already heard about. If there are missing transactions, the node fetches them from its peers. %The system is composed of two parts: a deterministic protocol (DECOR) to split rewards among all miners that generate a block of the same height, and a block propagation protocol (HOP) based on header-first propagation. 
%In this last protocol, the header is efficiently propagated using a push model (not requiring a round-trip), and the block data is propagated as in \bitcoin. 
%These enhancements allow \decorehop to achieve a throughput of 30 transactions per second and a block frequency of 60 blocks per second.

% ======

%\sarahm{Need to say in the conclusions that we don't really talk a lot about synchrony, or say in the methodology that our evaluation framework doesn't capture it (we always assume weak synchrony or something).}
%\subsection{The Zeitgeist Attack} 
%\alberto{if we really don't have enough space this section can go}
%\Nakamoto consensus assumes a synchronized clock across peers, as timestamps play a major role in calculating the target value $t$ in \pow puzzles. Miners include the current timestamp in each block. \bitcoin nodes only accept blocks with timestamps that are a maximum of 2 hours ahead of the node's `network-adjusted time' (the median of the timestamps returned by all nodes connected to the node---which can be at most 70 minutes ahead of the node's local timestamp). This means that each a block's timestamp may have up to 3.17 hours of error. Theoretical attacks have been discussed in the \bitcoin community which take advantage of this inaccuracy~\cite{timestampattack1,timestampattack2} by allowing miners to lie about the timestamp in mined blocks to artificially lower the value of $t$, and consequently gain more reward than their hash rates would warrant, or lowering the inter-block interval.

\subsection{Incentives}
%\subsubsection{Selfish Mining} 

\label{pow:selfishMining}

The security of Nakamoto consensus relies on economically incentivising miners to validate and mine blocks, by rewarding them with new coins. However, previous work has shown that Nakamoto consensus is not completely incentive compatible~\cite{Luu:2015:DIC:2810103.2813659,eyal2015minersDilemma,bonneau2016buy}. %Loi Luu et al. presented the \textit{verifier's dilemma}~\cite{Luu:2015:DIC:2810103.2813659}, where miners are incentivised to build on top of blocks without validating them, to save time and resources. The \textit{miner's dilemma} presented by Eyal~\cite{eyal2015minersDilemma} shows that mining pools can be incentivised to sabotage other pools by joining them but never sharing its proofs of work, causing the pool's participants to earn less and thus making them unattractive to participants. Bonneau showed that it is possible for an attacker to bribe existing miners to conduct an attack, rather than mining themselves~\cite{bonneau2016buy}.

Aside from incentives, protocol-level attacks exist that lower the security threshold of \bitcoin below 51\%. Selfish mining~\cite{selfishmining} allows colluding miners to generate more valid blocks than their computing power would normally allow them to if they were following the standard protocol. In selfish mining, colluding miners withhold blocks that they have found, which allows them to maintain a lead over the rest of the network, who may waste their computational power on stale blocks. When the network is about to catch up with the colluding miners, the colluding miners release a portion of their withheld blocks to the network. Using this mining strategy, it is possible to conduct a 51\% attack against the network with as little as 25\% of the network's computing power.

Systems like \fruitchain~\cite{pass2017fruitchains} aim to mitigate selfish mining by using two independent mining processes on top of each other: in addition to the \pow to create blocks, \fruitchain requires an additional \pow to mine an new type of block, called `fruits'. Blockchain transactions are included into these fruits, and the fruits are included into the blocks created by the first mining process. This mechanism prevents selfish miners from dropping honest blocks from the blockchain by releasing their withhold blocks because eventually, an honest block will be created and will include back all the dropped fruits.
\section{Proof-of-X Consensus}
\label{pox}

One of the biggest criticisms of \bitcoin is that it is based on power-intensive \pow that has no external utility, and makes it prone to centralization (Section~\ref{pow:miningCentralization}).
 %(consuming more electricity than Iceland according to a report~\cite{iceland})
%, it is claimed that over 70\% of the mining hardware is manufactured by a single company called Bitmain~\cite{bitmain70}. 
These limitations of \pow motivated a new class of consensus protocols based on proof-of-X (\pox) that replace wasteful computations with useful work derived from alternative commonly accessible resources, or remove computational work altogether.
% Here, we focus on proof of stake protocols where winning the leadership election is integrated with minting new blocks.}, proof of capacity that relies on quickly retrieving long-term stored data and trusted hardware solutions called proof of elapsed time.

\subsection{Proof-of-Stake}

In \poxStake, participants vote on new blocks weighted by their in-band investment such as the amount of currency held in the blockchain. 
A number of recent systems have provably secure \poxStake protocols~\cite{ouroborospraos,ouroboros,snow-white}. 
A common theme in these systems is to 
%randomly select a subset of participants from those that have stakes in the blockchain. 
randomly elect a leader from among the \stakeholders, which then appends a block to the blockchain. Leader election may be public, that is the outcome is visible to all the participants~\cite{ouroboros,snow-white}. Alternatively, in a private election the participants use private information to check if they have been selected as the leader, which can be verified by all other participants using public information~\cite{ouroborospraos}.
Private leader election is resilient to \dos attacks because candidates privately check if they are elected before revealing it publicly in their blocks, at which point it is too late to \dos them.
A malicious leader can censor transactions during its epoch. But as leaders are re-elected sufficiently often, a subsequent leader will add the censored transaction to the blockchain (albeit some delay).
%Thus transactions cannot be censored but only delayed. 
%These papers discuss the incentive structure without usually much formalization.
%The private approach is the one used in Ouroboros Praos and Algorand.
%Often, a leader is selected among a committee that has been established previously.

In \ouroboros~\cite{ouroboros}, the participants (a random subset of all stakeholders) run a multiparty coin-tossing protocol to agree on a random seed. 
The participants then feed this seed to a pseudo-random function defined by the protocol, that elects the leader from among the participants in proportion to their stake. 
The same random seed is used to elect the next set of participants for the next epoch.
\ouroboros distributes rewards among all the participants regardless of whether or not they win the election.
%Moreover, they add a notion of \emph{slot endorsers} who will endorse the input to be included in the block and get reward for each of their endorsement.
%rewards the input endorsers with the inputs that they have contributed
% Incentives: , availability and transaction
% verification should be incentivized
% , fees can be
% collected from those that issue transactions to be included in the ledger which can then be transfered
% to the block issuers.
% In our setting, similarly, a reward can be given to the parties
% that are issuing blocks and endorsing inputs. The reward mechanism does not have to be block
% dependent
% In our setting, it is possible to collect all fees of transactions
% included in a sequence of blocks in a pool and then distribute that pool to all shareholders that
% participated during these slots
% ) it rewards elected
% committee members for just being committee members independently of whether they issued a block or not
%  it rewards the input endorsers with the inputs that they have contributed. (iii) it
% rewards entities for epoch j, after slot jR + 4k.
%
% Because all the participants agree on the unique leader, in Ouroboros forks are less likely than in traditional Nakamoto consensus and , thus it provides strong consistency.
% They moreover show that their protocol satisfies the common prefix property, thus forks, if they happen, can only be
% of limited length. Thus Ouboros satisfies strong consistency.

In \ouroborospraos~\cite{ouroborospraos} and \snowwhite~\cite{snow-white} participants independently determine if they have been elected. 
%\snowwhite selects participants for each epoch based on the previous state of the blockchain, who independently check if they have been elected as the leader.  
\snowwhite uses similar criteria for leader election as \bitcoin, that is finding a \preimage that produces a hash below some target. However, participants are limited to compute only one hash per time step (assuming access to a weakly synchronized clock) and the target takes into account each participant's amount of stake.
\snowwhite employs the incentive structure of Fruitchain~\cite{pass2017fruitchains}: payouts are distributed equally among fruits (Section~\ref{pow:selfishMining}).
%Incentives:  We will distribute all forms of payout,
% including mining rewards and transaction fees to fruits rather than blocks. Furthermore, every time
% payout is issued, it will be distributed equally among a recent segment of roughly .. fruits.
In \ouroborospraos, participants generate a random number using a verifiable random function (\vrf). 
If the random number is below a threshold, it indicates that the participant has been elected as the leader, who then broadcasts the block along with the associated proof generated by the \vrf to the network.
\ouroborospraos inherits the incentive structure of \ouroboros.

A challenge for \poxStake systems is to keep track of the changing stakes of the \stakeholders. 
\ouroboros requires that shift in stakes is bounded, meaning the statistical distance is limited over a certain number of epochs.
Additionally, \snowwhite looks at stakes sufficiently far back in time to ensure that everyone has agreed on the stake distribution.

Outside academia, some deployed cryptocurrencies incorporate \poxStake~\cite{peercoin,blackcoin}, but their designs have not been rigorously studied and they are not very popular. Ethereum Foundation has been considering using \poxStake for some time, but their work is still in progress~\cite{casper}.

\subsubsection{Attacks and Mitigation}
\pox results in three new attacks compared to Nakamoto consensus~\cite{DBLP:journals/corr/Chepurnoy16}.
The first is called the \emph{nothing-at-stake attack} where miners are incentivized to extend every potential fork.
Since it is computationally cheap to extend a chain, in the case of forks rational miners mine on top of every chain to increase the likelihood of getting their block in the right chain.
One way of dealing with this is to introduce a penalty mechanism: a miner producing blocks on different forks is penalized by having part of their stake taken~\cite{snow-white}.
Another mitigation is to remove forks, at the cost of a bigger overhead~\cite{algorand}.
The second attack is called the \emph{grinding attack} where a miner re-creates a block multiple times until it is likely that the miner can create a second block shortly afterwards. This attack can be thwarted by ensuring that a miner is not able to influence the next leader election by using an unbiasable source of randomness or a deterministic leader election.
%\paddy{Paros has a "leaky resettable beacon" where the adv can see the beacon (i.e. leak) within a bounded time before others, can force the beacon to reset within a bounded time as well.}
In the third attack called the \emph{long-range attack}, an attacker can bribe miners to sell their private keys.
If these keys had considerable value in the past, then the adversary can mine previous blocks and re-write the entire history of the blockchain.
This is possible because the bribed miners have already received their external utility for these coins (\ie sold the coins for fiat currency), and no longer have a stake in the system. 
 Thus the bribed miners can send their keys to the adversary at almost no cost. This can be thwarted by central checkpointing: some entity (\eg one of the main developers) declares that some blocks are final if they are sufficiently far in time, or by requiring participants to lock their coins for a longer period of time than the duration of their participation.

\subsubsection{Alternatives}

Bonneau~\etal~\cite{bonneau:SOKbitcoin} describe informal (and unpublished) consensus protocols based on proof-of-stake that have been proposed in the cryptocurrency community. 
Broadly, these system require miners to hold or prove the ownership of coins.
We list three variations of this theme, though we note that this area has not seen significant advances.

\begin{itemize}[leftmargin=*]

\item \emph{Proof-of-deposit:} Miners `lock' a certain amount of coins, which they cannot spend for the duration of their mining. One such system is Tendermint~\cite{tendermint}, where a miner's voting power is proportional to the amount of coins they have locked.

\item \emph{Proof-of-burn:} Miners prove that they have destroyed a quantity of coins, for example by sending them to a verifiably unspendable address~\cite{proofofburn}. Slimcode~\cite{slimcoin} implemented this approach in 2014 but has recently been discontinued. 
%\paddy{bad internet but link is here https://bitcointalk.org/index.php?topic=613213.0}

\item \emph{Proof-of-coin-age:} Miners show possession of a quantity of coins, where the quantity of coins is weighted by their \emph{coin-age}---the time since the coins were last moved. Peercoin~\cite{peercoin} adapts this approach.

\end{itemize}

% ======

\subsection{Proof-of-Capacity}

In \poxCapacity, participants vote on new blocks weighted by their capacity to allocate a non-trivial amount of disk space. 
%Th aim is to require miners to be financially invested in disk-space and reduce the computational overhead of continuous mining.
%Note if mining is computationally cheap then this can also re-introduce the nothing-at-stake problem as previously discussed in proof of stake protocols.
%Also, there are intrinsic issues that must be overcome such as miners re-using storage for multiple sybils or miners outsourcing  storage an external provider.
\permacoin~\cite{permacoin} repurposes \bitcoin's \pow with a more broadly useful task: providing a robust, distributed storage.
In \permacoin, eligibility for the leader election requires participants to also store segments of a large file.
The file is distributed by an authoritative `dealer' who signs file blocks.
To provides censorship-resistant file storage, the file is fully recoverable from the participants in the event of a dealer failure or shutdown.
%Like \bitcoin \pow involves finding a \preimage that produces a hash that satisfies the networks difficulty, however the \preimage also includes a segment of the file. The \pow can be verified by replaying the \pow and replacing signing operations with signature verification.
\spacemint~\cite{Fuchsbauer2015SpaceMintAC} employs a consensus protocol based on a non-interactive variant of \poxCapacity (called \poxSpace), where participants generate and commit to a unique hard-to-pebble graph.
%In the original \poxSpace protocol, participants were required to respond to a challenge within time $T$ which is set such that there is not enough time to re-compute the graph. 
%\spacemint's non-interactive protocol relies on other miners publishing punishment transactions (similar to \bitcoinng's poison transaction) with proof that another miner's proof of space was incorrect.
%This punishment transaction must be received into the blockchain within a bounded time $T$, otherwise the incorrect proof is accepted.  
\permacoin and \spacemint have the same basic model as Nakamoto consensus, so inherit \bitcoin's incentivization mechanism, as well as its resilience against censorship and \dos.

\subsubsection{Attacks and Mitigation}
Proof-of-capacity is vulnerable to centralization due to participants outsourcing the file storage to an external provider. To mitigate this problem, the \poxRetrieve in \permacoin requires sequential read access to blocks in a pseudorandom order: this directly increases the bandwidth latency in case of outsourced storage, which reduces the miner's chance of finding a solution.

\subsection{Proof-of-Elapsed-Time} 

Using the trusted enclave in Intel SGX, it is possible to replace computational work with \poxElapsed~\cite{sawtooth}.  Participants request a wait time from their enclave and the chip with the shortest wait time is elected as the leader.  The newly elected leader can provide an attestation alongside the new block to convince other participants that: \first it indeed had the shortest wait time, and \second that it did not broadcast the block until after the wait time had expired. 

An alternative approach is called Resource-Efficient Mining (REM)~\cite{rem} that proposes computing useful \pow using trusted hardware. 
Every instruction cycle for the useful \pow can be seen as a lottery ticket: if a cycle wins the lottery, the participant is authorized to mint a new block.  
To extend this model to arbitrary work, the authors introduce a two-layer hierarchical attestation.
The first layer certifies that useful \pow was performed, and the second layer attests that the program (and its input) incremented the counter for instruction cycles appropriately. 
A hash of both layers is sent alongside a new block to prove that the participant was authorized to mint it. % REM can detect SGX that mint strangly frequently block (improbabale)

\subsubsection{Attacks and Mitigation}
Both \poxElapsed approaches suffer from two limitations. 
First, breaking a single piece of trusted hardware enables the attacker to always win the lottery. 
Both Sawtooth and REM argue that a statistical analysis of newly minted blocks suffices to detect whether a chip can be compromised. 
Second, the \emph{stale chip problem} highlights that it is advantageous to collect chips as this increases the probability of minting a new block (\ie every new chips is an additional lottery ticket). 
REM provides an economic analysis to show that a miner's revenue source originates from useful work, and not farming chips.  
\section{Hybrid Consensus: Single Committee}

\label{hybrid-single}

A single consensus node suffers from poor performance as well as safety limitations such as weak consistency and low fault-tolerance. This has resulted in a shift towards consensus protocols where  a \emph{committee}---rather than a single node---collectively drives the consensus.  

%The committee can tolerate up to $1/3$ byzantine nodes, and is more resilient to \ddos as it is more challenging to simultaneously take down all honest nodes in the committee. The main advantage of this approach is strong consistency and instantaneous confirmation. However, \bft protocols do not scale well and are not suitable for decentralized networks. To address these issues, sharding is being explored where the consensus protocol is run within and across multiple smaller committees (Section~\ref{multicommittee}). 

% ============

\subsection{Committee Formation}

\label{bootstrapCommittees}

Committee formation refers to the criteria used to allow nodes to join a committee. This is an important aspect of decentralized, permissionless systems to thwart \sybil attacks.

\subsubsection{Permissioned} 

Permissioned blockchains operate in a trusted environment where nodes are granted committee membership based on organizational policy. \hyperledger~\cite{cachin2016architecture} is one such system that supports smart contracts. There is a hierarchy of trust, with some nodes being fully trusted while others only partially trusted. This allows for a modular design where transaction validation is performed by the fully trusted nodes (or endorsers) while the semi-trusted nodes (ordering nodes) order the transactions and add these to the blockchain. In \hyperledger, clients first submit their transactions to the endorsers who execute the smart contract. A transaction is only submitted to a subset of endorsers according to the policy of the respective smart contract. As different smart contracts can designate different endorsers, execution can take place in parallel. Clients collect matching signed results and smart contract state updates from sufficient number of endorsers, and submit these to the ordering nodes which append it to the blockchain using a consensus protocol.
% \bano{This doesn't seem interesting. George already mentions this in discussion on exploiting advances in hardware and crypto for better consensus protocols. I am commenting it.}
%Abraham~\etal~\cite{abraham2016bvp} describe a framework called Byzantine Vertical Paxos (BVP) aiming to achieve high throughput \bft state machine replication. Extending the work of~\cite{kotla2007zyzzyva, van2012byzantine}, BVP builds a steady state protocol and a reconfiguration protocol on top of Vertical Paxos. %Reconfiguration is achieved through a wedging scheme, where a coordinator obtains the latest state of the system and is validated from a subset of nodes. The coordinator is then responsible of driving a reconfiguration consensus decision implemented by a separate Byzantine consensus. 
%If the network is assumed to be synchronous, BVP requires only $f+1$ nodes for the stead state protocol and $2f+1$ nodes for reconfiguration. On the other hand, if working with an asynchronous network, BVP provides a schemes relying on Trusted Platform Module.

% ------

\subsubsection{Proof-of-work} In these systems, nodes are allowed to join the committee based on \pow. 
In \byzcoin, the consensus committee is dynamically formed by a window of recent miners. Each miner has voting power proportional to its number of mining blocks in the current window, which is proportional to its hash power. When a miner finds a solution to the puzzle, it becomes a member of the committee and receives a share in the consensus. 
\solidus and \omniledger have a similar model for committee formation. 
\omniledger also supports \poxStake to allocate committee membership based on directly invested stake instead of power-wasteful work. A public randomness or cryptographic sortition protocol is run within the current committee to select the next committee from the current stakeholder distribution defined in the ledger. 

% -------

\subsubsection{Lottery} Candidates are promoted to committee membership based on the outcome of a lottery. In \algorand, all candidates have a public key, and get chosen to become a committee member using cryptographic sortition. This involves the candidates running a verifiable random function and seeing if the output is below a certain value. 

% ============

\subsection{Committee Configuration}

The way a committee is configured has safety and performance implications. Permissioned systems usually assume static committee members, but \sybil resistance in a permissionless and decentralized setting requires dynamic membership. 

\subsubsection{Static} In static setting, the committee members are not periodically changed. This is the typical configuration in permissioned systems like \hyperledger and \rscoin where committee members have known, trusted identities and the threat model does not include \sybil attacks.

% ------

\subsubsection{Rolling (Single)}
The committee is updated in a sliding window fashion: new miner(s) are added to the current committee and the oldest members are ejected. In \byzcoin, each miner has voting power proportional to the number of mining blocks it has in the current window, which is proportional to its hash power. When a miner finds a solution to the puzzle, it becomes a member of the current consensus group and receives a share in the current window which moves one step forwards (ejecting the oldest miner). 

An important aspect of reconfiguration is \emph{wedging}, that is to stop the old committee from approving more transactions without losing any transactions it is processing at the time of reconfiguration. \solidus updates its committee similarly to \byzcoin, but a new miner joining the committee can propose transactions only once. This binds transaction proposals to reconfiguration, so it is no longer possible for an old committee to approve transactions concurrent to a reconfiguration event. 
Another issue is how to resolve \emph{leader contention} when two miners simultaneously solve a \pow puzzle. \solidus uses a Paxos-style leader election where a higher ranked leader can interrupt a lower ranked leader. Ranks are derived from leaders' \pow solutions and supplementary epoch numbers. To ensure safety, the new leader must propose a value that has been (or may be) committed.   
%The new leader selects a rank r that is higher than the previous leader and sends a new-leader message to  the committee. Committee members verify that the previous leader has expired and that the new leader should be the next leader. If this check passes, then the committee responds with the (a) value accepted from the previous highest ranked leader, (b) or if no value was accepted then it responds with a promise that no proposals with ranks lower than r will be accepted. After receiving responses from 2f+1 committee members, the leader proposes a safe value with a proof attached: a safe value is one proposed by the highest ranked leader among those accepted by the 2f + 1 committee members.
%According to a follow-up blog post (\texttt{http://hackingdistributed.com/2016/08/04/byzcoin}), \byzcoin can deadlock and permanently lose liveness during reconfiguration. Most existing systems other than \solidus do not discuss wedging and reconfiguration. 

A reconfigurable committee needs some mechanism to track committee membership. 
%It seems that only \peercensus discusses this. I also recall \smartbft supports changes in the 'view' (dynamic membership).
In \peercensus~\cite{decker2016bitcoin}, a new member is allowed to join the committee following a collective decision which involves validating that the member is reachable over the network. Committee members use a failure detector (e.g., by sending regular ping messages) to detect when a member has left the committee. If a member finds another to be unreachable, it can propose `leave' for the absent member and the committee membership is updated after a collective decision is made by the committee. A limitation of this approach is that malicious members can slow down or stall the system by constantly generating false alarms for eviction of legitimate members. Addressing this would require rate-limiting the number of leave operations a member can propose in a given time interval.
%\elastico counts failed nodes as the f malicious fraction.
%``We discuss only protocols for static groups here; they require explicit group reconfigura- tion [67, 9] and do not change membership otherwise. This assumption contrasts with view-synchronous replication [26], where the group composition may change implicitly by removing nodes perceived as unavailable.''~From~\cite{cachin2017blockchains}

% ------

\subsubsection{Full} Lottery-based systems like \algorand and \snowwhite select the committee members for each epoch using randomness generated based on previous blocks.

% ------

\subsubsection{Rolling (Multiple)}
\omniledger uses cryptographic sortition to select a subset of the committee to be swapped out and replaced with new members. This is done in such a way that the ratio between honest and byzantine members in a committee is maintained. This also has the benefit that the system is operational during reconfiguration as the operational members can continue to process transactions while a fraction of the committee is being reconfigured and bootstrapped.     

% =============

\subsection{Consensus Protocol}

Most committee-based systems use classical \bft consensus protocols such as \pbft. In this section we focus on modifications to classical \bft protocols or their novel compositions to tailor them for use in blockchains.   

In \solidus, the leader is external to the committee and can propose transactions and \pow to nominate itself as a committee member only once to the committee. If the committee agrees, they approve the proposed transactions and allow the miner to join the committee in the next round. The proposal, that has now become a decision, also serves as the next puzzle and is propagated to all miners. This approach is motivated by a safety problem in \pbft's `stable' leader which can potentially manipulate reconfiguration by waiting for a malicious miner to solve the puzzle, and later nominating it on to the committee---allowing the committee to gradually become dominated by corrupt members. 

 \byzcoin organizes the consensus committee into a communication tree where the most recent miner (the leader) is at the root. The leader runs \pbft~\cite{pbft} to get all members to agree on the next block. However, it replaces \pbft's $O(n^2)$ MAC-authenticated all-to-all communication with a primitive called scalable collective signing (\cosi) that reduces messaging complexity to $O(n)$. The outcome of running two rounds of \pbft with \cosi is a fixed 64 byte collective signature that proves that at least two-thirds of the committee members witnessed and attested the block. A node in the network can verify in $O(1)$ time that a block has been validated.

A malicious committee leader can potentially censor transactions by not proposing them; this does not compromise safety, but it can negatively affect fairness. \omniledger deals with this issue by allowing non-leader committee members to propose a transaction if they suspect that it has been censored by the leader (since they can `hear' all messages because of the gossip protocol for information dissemination). In \chainspace and \byzcoin, transaction-censorship triggers leader re-election (or view change). \elastico does not discuss censorship by committee leader. Classic BFT protocols often rely on a timeout to detect censorship or unreliability by a leader, and thus liveness and censorship resistance rest on a partial synchronous network assumption. % \bano{George: You once mentioned that timeout-based approaches can be problematic. State here.} 

In the context of \bft-based committees, \dos means that all the honest members in the committee are taken offline. This is challenging, and made further difficult by how frequently and what fraction of committee membership changes (epoch, dynamism). \byzcoin has medium \dos resistance because the committee configuration is rolling (single). \elastico has high resistance within single committees because full committees are reconfigured (full swap), but in the absence of an intra-shard consensus mechanism, an adversary can flood the system with transactions that touch multiple committees causing a deadlock. \omniledger further enhances \byzcoin's efficient \bft protocol with $O(n)$ messaging complexity by using a more robust group communication pattern. This addresses an issue in the original protocol where node failures cause \byzcoin to fall back on a more robust all-to-all communication pattern that significantly degrades system performance. \chainspace does not include details on intra-committee configuration, hence it only provides medium level of resistance against \dos attacks.   

Systems based on \poxStake consensus (\eg \ouroborospraos and \algorand) achieve \dos protection by privately electing committees. This ensures that
participants cannot learn whether another participant is a committee member and only learns this when the newly elected member announces this.
In Algorand, users check for themselves whether or not they should play a role in the committee for the next round by seeing if, on input a seed known only to the user, the output of a verifiable random function is less than a certain value; this ensures that only they know what roles (if any) they should play in the committee.  Once the roles are fixed, users then participate according to their role in the \textit{BA}$\star$ consensus protocol, which is a variant of PBFT that allows the set of participating servers to rotate.  As participants start playing their roles, they can include information in their messages that allows other participants to check that they are in fact eligible.
 
\hyperledger uses \emph{pluggable and modular} consensus in which the consensus protocol can be specified by the smart contract policy. For example, \hyperledger supports a \cft service based on Apache Kafka~\cite{apache-kafka} and its ZooKeeper unit~\cite{Hunt:2010:ZWC}, and more recently a \pbft-variant \bftsmart~\cite{bftsmart}. Because of their trust assumptions, permissioned systems are resilient to \dos and censorship of transactions by the committee.  
 
 % ==========
 
\subsection{Incentives}

Classical \bft protocols assume two kinds of players: cooperative and byzantine. This assumption works well in centralized settings where nodes are controlled by the same entity or federation. However, decentralized networks that rely on volunteer nodes need to provide incentives for participation.

Most committee-based systems such as \byzcoin use the same incentive model as \bitcoin; however, instead of the most recent miner receiving all reward and fee, it is shared between members of the committee in proportion to their shares. \byzcoin states that to ensure that members remain active after joining the committee, they will also be rewarded for participation (\eg upon completion of \pbft pre-prepare and commit phases); however, details have not been provided.

In general, consensus protocols assume two kind of players: honest and byzantine. \solidus argues that with no clear incentives, the honest (or altruistic) committee members have nothing to gain from participating in the consensus. To alleviate this, \solidus introduces a third kind of player: a \emph{rational} player that assesses its expected utility in terms of \solidus coins. \solidus argues that equal distribution of rewards between all committee members can lead to a situation where members can suppress reconfiguration to stay on the committee and continue to collect rewards.  To address this, \solidus rewards the committee members that are the fastest to endorse reconfiguration creating a competition. Moreover, the 
%reward from each committed block consists of a fixed mining reward and some varying transaction fees. The 
external leader gets all the transaction fees, while the mining reward is split between the external leader and the committee members. This ensures that committee members do not have an incentive to delay reconfiguration due to a possible high transaction fee.  
%The reward for committee members is further divided into four shares, one for each phase in the multi-round consensus protocol. 
\solidus includes incentives for information propagation and present a game-theoretic analysis that a miner’s best strategy is to propagate the \pow puzzle and charge a small fee. 
%Such incentives have also been discussed in \texttt{Moshe Babaioff, Shahar Dobzinski, Sigal Oren, and Aviv Zohar. On bitcoin and red balloons. In Proceedings of the 13th ACM conference on electronic commerce, pages 56–73. ACM, 2012.} 

Smart contract platforms require clients to include fees to be paid to the nodes that execute the smart contracts. This not only helps to incentivize node participation, but also protects the system from overuse by discouraging clients from submitting long computations that monopolize system resources. \ethereum clients have to pay `gas' in proportion to the cost of executing the contract~\cite{wood2014ethereum}.
\section{Hybrid Consensus: Multiple Committees}

\label{hybrid-multi}

While single-committee consensus significantly improves performance over single-node consensus, a major limitation is that it is not scalable: adding more members to the committee decreases throughput. This motivated the design of consensus based on multiple committees. To make the system scalable, transactions are split among multiple committees (shards) which then process these transactions in parallel.  

\subsection{Committee Topology}

When multiple committees are involved in consensus, an important question is how they will be organized in terms of topology. \chainspace and \omniledger have flat topologies, that is all committees are at the same level. \elastico has a hierarchical topology in which a number of `normal' committees validate transactions, and a leader committee orders these transactions and extends the blockchain. 
%In each round, nodes submit \pow to the leader committee which  uses the \pow’s least-significant bits to allocate the miners to different committees. Within committees, \pbft is run.  A committee is responsible for processing only a subset of transactions, but node reconfiguration takes place every round implying that each node has to save the global state which can hinder performance. 
In \rscoin~\cite{rscoin} (a permissioned blockchain), the central bank controls all monetary supply, while mintettes authorized by the bank validate a subset (shard) of transactions.
The transactions that pass validation are submitted to the central bank which adds them to the blockchain.

\subsection{Intra-committee Configuration}

In permissioned systems, the process of assigning nodes to committees is usually done statically according to the policy of the federation. Another approach is to dynamically allocate nodes to committees. This should be done randomly to stop an adversary from concentrating its presence in one committee and exceeding the byzantine-tolerance threshold. Permissioned systems like \rscoin can use a trusted source of randomness for committee reconfiguration, but this can be problematic in a permissionless setting which would require a shared random coin~\cite{Corbett:2013,Glendenning:2011}. However, generating good randomness in a distributed way is a known hard problem: current best solutions tolerate up to $1/6$ fraction of byzantine peers, while incurring a high message complexity~\cite{Awerbuch:2006}. Among the more recent solutions, \randhound~\cite{randherd} provides a scalable, secure multi-party computation protocol that offers unbiasable, decentralized randomness in a byzantine setting. 

\omniledger periodically reconfigures committees to  ensure that a committee is never compromised. This is achieved by a secure shard reconfiguration protocol, based on \randhound, that committee members run periodically and  autonomously. In every epoch, a random subset of members is replaced with new set of members that registered their interest in the previous epoch. The swap operation is done such that liveness is maintained during reconfiguration events because a subset of committee members continues to be operational.  

\elastico operates in epochs: assignment of nodes to committees is valid only for duration of the epoch. At the end of the epoch, nodes compute solution to a puzzle seeded by a random string generated by the final committee and sends the solution to the final committee to be assigned to a committee. As a result, in each epoch a node is paired with different nodes in a committee managing a different set of transactions. The number of committees scales linearly in the amount of computational power available in the system, but the number of nodes within a committee is fixed. 
%Thus the block throughput scales up almost linear to the size of the network. As more nodes join the network, transaction throughput increases without adding to latency as messages needed for consensus are decoupled from computation and broadcast of final block to be added to the blockchain.

\chainspace has abstracted details of committee reconfiguration and it is up to policy enforced \via a smart contract to decide how nodes will be allocated to committees. Nodes can be added (and removed) to committees by their members through majority ($2f+1$) voting.

% \bano{Already included in Discussion}
%Most systems do not discuss how members of the same committee discover each other following reconfiguration, the underlying assumption being that this can be achieved \via a gossip protocol. \elastico has an explicit overlay setup (a fully-connected subgraph) for committees that describes how members in the same committee will discover each other. Instead of broadcasting information which has $O(n^2)$ messaging complexity, they provide a methodology that requires $O(nc)$ broadcast messages, where c is the number of committees. A special committee serves as a set of directories which can be queried by a new member to find other members in its committee. The directories and the committee members can tolerate different views of the member set up to a threshold.

% /bano{Dropping this. Not strong enough.}
%\subsubsection{Transaction-to-Committee Mapping}
%Dynamic: Non-voting peers in \bigchaindb assign transactions randomly with equal probability to committees.

\subsection{Intra-committee Consensus}

In a multi-committee system, some transactions might involve coordination between multiple committees. Such transactions might require access and manipulation of state that is handled by different committees. The intra-committee consensus ensures that this takes place consistently and atomically across all concerned committees. 
%\elastico does not provide transaction atomicity across shards.

%\omniledger uses a block-DAG (Directed Acyclic Graph) rather than a blockchain, effectively creating multiple blockchains in which consensus of transactions can take place in parallel. To realize parallel consensus, dependencies between transactions are identified from their inputs and outputs. Moreover, transactions are organized in such a way that the block containing a transaction must be a member of the blockchain corresponding to the transaction's inputs. 
 \omniledger uses an atomic commit protocol to process transactions across committees. A transaction submitted by a client is processed by the committees that manage its inputs. Each related committee validates the transaction, and returns a proof-of-acceptance (or rejection) to the client, and locks the transaction inputs. To unlock the inputs, the client sends proof-of-accepts to the committees that manage the transaction outputs, who add the transaction to the next block to be appended. If the transaction fails the validation test, the client can send proof-of-rejection to the input committees to roll back the transaction and unlock the inputs. 
  
In \rscoin, communication between committee members takes place indirectly through the client (similar to \omniledger), and it also relies on the client to ensure completion of transactions. 
%The central bank controls all monetary supply, while committees of mintettes (nodes authorized by the bank) manage and validate a subset of transactions assigned to them. 
%Transactions have identifiers, and each mintette is responsible for a subset (shard) of transactions such that a shard can potentially overlap across mintettes for security and reliability. A mintette maintains information about outputs of the transactions it manages, whether these have been spent and if so in which transactions. 
A client first gets signed `clearance' from majority of the mintettes that manage the transaction inputs. Next the client sends the transaction and signed clearance to mintettes corresponding to transaction outputs. The mintettes check validity of the transactions and verify signed evidence from input mintettes that the transaction is not double-spending any inputs. If the checks pass, the mintettes adds the transaction to be included in the next block.
%send evidence to the client that the transaction will be included in the blockchain (this evidence can be used to implicate the mintettes if they misbehave). 
The system operates in epochs: at the end of each epoch, mintettes send all cleared transactions to the central bank which collates transactions into blocks that are added to the blockchain. 
%As communication between mintettes takes place indirectly through the user, \rscoin has low communication overhead and improved performance. The transaction throughput scales linearly with the number of mintettes.

Client-driven atomic commit protocols like \omniledger and \rscoin are vulnerable to \dos if the client stops participating and the inputs are locked forever. These systems make the assumption that clients are incentivized to proceed to the unlock phase. Such incentives may exist in a cryptocurrency application where an unresponsive client will lose its own coins if the inputs are permanently locked, but do not hold for a general-purpose platform where transaction inputs may have shared ownership. Instead of a client-driven approach, \chainspace runs an atomic commit protocol collaboratively between all the concerned committees. This is achieved by making the entire committees act as resource managers for the transactions they manage. 

% \bano{Already included in the Discussion section.}
%A limitation of existing multi-committee systems is how to tolerate malicious shards. None of the existing systems can tolerate even a single malicious committee. While auditability mechanisms have been discussed in the context of byzantine members within a committee, there are no mechanisms to even detect a corrupt committee. \bano{Needs polishing}

%\elastico does not talk about the possibility of the final committee censoring transactions.
%Both \omniledger and \chainspace do not explicitly state the messaging complexity of the intra-committee consensus protocol. 

% \bano{Interleave this discussion in relevant sections in hybrid systems}
%\subsection{Performance}
%In multiple-committee systems, latency involves the time to reconfigure committees (if dynamic), the time to perform inter-committee consensus, and the time to perform intra-committee consensus.  In \omniledger, it is hard to separately evaluate latency due to the intra-shard consensus protocol. This cost is included in the reported number for latency. 

% =================
\section{Discussion}

\label{discussion}

\subsection{Integrating \bft protocols into blockchains}

The renewed interest in BFT protocols, in the context of blockchain, has led to more mature and efficient variants of those protocols, or variants that leverage new assumptions. Here we discuss several open problems that still remain.

\subsubsection{`Open' vs `closed' asynchronous protocols}
%\bano{TODO: George}

A number of recent scalable blockchain protocols, such as RSCoin~\cite{rscoin}, Omniledger~\cite{omniledger} and Chainspace~\cite{chainspace}, employ traditional byzantine consensus protocols for scalability and sharding. However, those consensus protocols are inherently `closed', in the sense that replicas need to have authenticated channels between them, long term interactions with each other, and can only tolerate $f$ byzantine nodes. Thus, traditional consensus protocols cannot accommodate open participation of nodes and high churn, and are vulnerable to \sybil attacks~\cite{douceur2002sybil}. 

Newer BFT protocols, such as Honeybadger~\cite{miller2016honey}, even overcome impossibility results, and provide both safety and liveness in a fully asynchronous setting, through a randomized consensus algorithm. While this breakthrough, building upon the earlier work by Cachin~\etal~\cite{cachin2000random} is of notable theoretical value, it does not resolve the issue of the need for a `closed' group and therefore those solutions cannot be a drop-in replacement for open Nakamoto consensus. Such randomized BFT protocols have traditionally been more expensive than deterministic ones, both in terms of communication and cryptographic operation costs.
Byzantine consensus protocols, besides Nakamoto consensus, in the context of open group participation is still an open research problem.

\subsubsection{Exploiting advances in hardware and cryptography}

The most mature current implementation of BFT is the Java \bftsmart~\cite{bftsmart} library with message complexity $\mathcal{O}(N^2)$ in the size of the quorum $N$. However, Byzcoin~\cite{byzcoin} uses modern signature schemes to optimistically relay all messages through a leader, reducing the common case of BFT consensus to $\mathcal{O}(N)$. The XFT~\cite{liu2016xft} protocol, on the other hand, improves the efficiency of consensus by relaxing the threat model. It considers that byzantine nodes may act arbitrarily, however links between honest nodes are reliable and eventually synchronous. This leads to a simplification of the view change and steady state BFT protocol. Finally, some consensus protocols are now leveraging secure hardware executions environments: the Intel Sawtooth lake system uses the Intel SGX and related trusted execution environments to perform the duties related to ordering transactions, while ensuring safety and liveness~\cite{prisco2016intel}. 
%Those novel solutions to traditional BFT consensus protocols, have been catalyzed by the need of blockchain protocols, and will likely be a lasting legacy of those systems.

\subsubsection{Identity management}

%\BFT consensus protocols involve all $n$ committee members of which $f$ can be byzantine. The protocol proceeds in rounds, and the outcome of each round depends on responses from $f+1$ members (assuming at least 1 honest member). 
In \bft consensus protocols, a malicious member can potentially generate spoofed responses on behalf of other members to bias majority in its favour. To counter this attack, \bft committees assume that there exist point-to-point, authenticated channels between all members, which requires some mechanism to track committee members and their keys. Tracking membership and key distribution in dynamic permissionless committees is challenging, and most systems abstract these details. 

A na\"ive solution is for all nodes to regularly broadcast their identity to the entire network (along with evidence that they have been granted permission to join the committee) resulting in $O(n^2)$ messages. A better approach to is to form a special committee that offers directory services to new committee members~\cite{elastico}. However, this presents a dilemma: a static committee undermines decentralization, but forming a decentralized directory committee suffers from the same challenges as the committee aims to solve in the first place. Another technique, used by \omniledger~\cite{omniledger}, is to record committee members for each round in a separate `identity' blockchain---however, its details are not provided. In multi-committee systems, intra-committee interaction further requires each committee to have a collective identity, and some way for the committees to discover each other. 
%
%\subsubsection{\BFT libraries} It is widely agreed that consensus protocols are complex and hard to implement~\cite{Chandra:2007,Guerraoui:2010}. While these protocols have been actively studied for well over two decades, only prototypes developed in academic papers to validate findings are available and complete implementations are hard to find. \bftsmart is the most robust and well-known open-source Java library that implements \bft state machine replication~\cite{bftsmart}. The authors note that they started developing it in 2010 and took five years to complete. \bftsmart has been understandably evaluated in a modest network of 4 machines in a LAN as it was developed before the interest of the blockchain community in \bft protocols surged. For this reason \bftsmart is not suitable to be directly plugged into blockchains: the library should be re-evaluated in such settings and the software should evolve accordingly. Recently the authors of \bftsmart combined it with their extensions to support low-latency in geo-replicated machines~\cite{Sousa2015Wheat} and incorporate it in a popular permissioned blockchain, Hyperledger~\cite{Sousa2017hyperledger}. (The code is available online~\cite{hyperledger-bftsmart}.) Given this context, the rate at which a plethora of mostly esoteric blockchain consensus protocols have been (and continue to be) developed over a short period of time is worrying. 

\subsection{Committee-based approaches} 

\subsubsection{Secure committees}

The idea of scaling services built on state machine replication (\smr) by splitting state (or sharding) among multiple committees (also called partitions or shards) has been well-studied in the context of traditional distributed systems~\cite{Corbett:2013,Glendenning:2011,Le:2016:SSMR}. The key challenge in these systems is to ensure linearizability by atomically executing operations that span multiple committees. More recent optimizations enable \emph{elastic} \smr so that committees can dynamically merge (scale up) and split (scale out) their state for load-balancing purposes~\cite{NogueiraCB17}. These systems employ fault-tolerant \bft protocols at their core as the nodes are controlled by a single entity or a group of entities that collectively govern the system. Due to similar governance assumptions, these techniques can be extended to permissioned blockchains. However, sharding permissionless blockchains with byzantine adversaries is challenging and tackled by only a few recent systems~\cite{chainspace,omniledger,elastico}. Individual committees can tolerate up to 33\% of malicious members, but if this is not the case then the malicious committee can compromise all the transactions that touch the bad committee. This is an outstanding issue shared by all multi-committee blockchains. Future research should focus on developing robust mechanisms to detect malicious committees and to recover from them.

\chainspace starts mitigating this issue by making the author of the smart contract responsible to designate the parts of the infrastructure that are trusted to maintain the integrity of its contract; the contract's integrity only depend on their correctness (as well as the correctness of contract sub-calls). Moreover, \chainspace provides an auditing mechanism allowing honest node in honest committees to detect inconsistencies and discover the malicious committee; there are however no systems today providing a recovery mechanism.

Finally, sharded solutions achieve a different notion of verifiability from solutions that rely on a single committee (or are fully decentralized), as it is no longer clear how to define a global set of transactions.  In \omniledger and \chainspace, for example, every committee defines its own blockchain, and
in \rscoin the separate sets of transactions agreed upon by each shard are combined only through the use of a central entity.  We leave it as an interesting research problem to quantify the difference, in terms of public verifiability, between sharded and non-sharded solutions.

\subsubsection{Bootstrapping committees} 

The biggest threat to the integrity of a permissionless committee is from an adversary that might create \Sybil identities and take over the whole committee. As discussed in Section~\ref{bootstrapCommittees}, prominent approaches include using \pow or \pox to allow nodes to join the committee. A limitation here is that the biggest miners will have a greater likelihood of dominating the committee, though at the cost of significantly more hashing power than required for single-leader \pow systems. Other \pox alternatives have been proposed but these suffer from similar issues. 

Multi-committee systems raise the additional issue of how to map nodes to committees. One approach is to randomly map nodes to committees~\cite{omniledger,elastico}. However, this prohibits finer governance. General-purpose platforms like \chainspace might have different policies within committees; for example some committees can be permissioned while others are permissionless. In this case it might be useful to enforce node-to-shard mapping via smart contracts that allow a node to join a committee trusted by the smart contract provider. 

Another consideration for bootstrapping committees is to achieve coercion resistance, in the form of requiring enormous effort for an adversary to suppress the overall operation of the system. Systems such as Tor~\cite{DBLP:conf/uss/DingledineMS04} have survived in a highly adversarial environment despite parts of its infrastructure, namely directory authorities, being a closed consensus group. These authorities are distributed geographically, and are under different jurisdictions and managed by different organizations. %to preclude both collusion and single jurisdiction attacks. 
Furthermore, like blockchains, they only handle high-integrity operations---not privacy-sensitive ones---making their audit and also replacement in case of unreliability, easier despite being manual. This is a hopeful example, illustrating that even small closed groups may, through careful selection of participants, provide sufficient protection against coercion.

\subsection{Incentives and governance in consensus protocols}

Decentralized networks need to incentivize nodes for active participation in different operations such as consensus~\cite{byzcoin,elastico,omniledger}, information propagation~\cite{babaioff2012red,solidus,meshcash}, and executing smart contracts~\cite{chainspace,wood2014ethereum}.
Recently there has been a shift towards incentive-compatible consensus protocols, where incentives are built into the core of the protocol. \solidus argues that cooperative players in the classical \bft protocols are replaced by rational players in the decentralized setting, who are motivated to maximize their utility~\cite{solidus}. To ensure that rational players participate in all phases of the protocol, incentives should be distributed among them such that they can only be claimed after the completion of each phase. This is a new but important observation. This implies that the committee should be reconfigured regularly to maintain a suitable committee size: large committees might result in trivial rewards for individual committee members (or might lead to inflation of client fees to account for the difference). 

An important question is: who distributes the incentives? In \solidus the leader of the committee distributes incentives among the first $2f+1$ responders. This approach has several limitations: \first a faulty or malicious leader might not divide the rewards, \second there is no way to enforce that the leader rewards the genuinely fast responders, so the leader can instead wait for its favourite members to reply, and \third the notion of `fast' is problematic in decentralized networks where members located farther from the leader are at a natural disadvantage.

Broadly, incentivization in \pow blockchains has seen some study~\cite{kroll:2013}: major limitations have been identified~\cite{sapirshtein2015optimal,selfishmining,nayak2015stubborn,Carlsten:2016:IBW:2976749.2978408}, and possible solutions have been proposed~\cite{zhang2017publish}.  Similarly, in the context of protocols where creating a block is cheap, good incentives are crucial to prevent attacks on the system, but have not been carefully analyzed.  
The investigation of these issues in \bft protocols is likewise far from mature, and non-existent in multi-committee protocols where the incentives need to be extended to intra-committee operations. This area will benefit from combining formal economic and game theoretic analysis with cryptography, such as has already been done in the blockchain community~\cite{poon2015bitcoin,smartcast}.  Techniques such as rational cryptography~\cite{rationalcrypto,DBLP:journals/corr/abs-1005-0082} and the BAR model~\cite{barmodel}, which considers Byzantine, altruistic, and rational agents, could also be adapted to work here.

Beyond concrete incentivization to participate in the protocol, it is important to consider also what makes protocols attractive to participants in the first place, and what makes them think their investment in a given system  will be repaid.  These broader types of incentive are both related to the governance of the system, in terms of identifying the entities who define its rules, and the extent to which the protocol is able to evolve.
%
%With respect to the rules, even the most decentralized consensus protocols are heavily centralized, with the various parameters (block reward size,  sblock rate, etc.) being defined by one person or small set of people and remaining largely unchanged since.  In the Zcash cryptocurrency, for example, the governing entities introduced an explicit ``founder's reward'' that pays 5\% of all generated coins back to its creators.
%
%With respect to protocol evolution, different systems have taken radically different approaches.  In Bitcoin, for example, a debate over the size of blocks has already led to one hard fork of the system (into Bitcoin and Bitcoin Cash) and is in the process of leading to another at the time of writing, with the main governing entities bitterly divided over which route to take.  In Ethereum, in contrast, a more commanding set of governing entities has overseen at least five hard forks (with only one, after The DAO attack, leading to a rival cryptocurrency in the form of Ethereum Classic), with many more promised in the future.
%
It is ultimately a relatively unstudied question at this point what types of governance structures would provide the strongest incentivization, or the extent to which these structures are taken into account when participants are deciding which protocol to join.  This area would benefit from social science-based analysis.

%\subsection{Provable Guarantees}
%
%Traditionally, designing and implementing consensus protocols has been deemed a complex undertaking: the classical protocols were accompanied by formal proofs that in some cases spanned entire PhD dissertations~\cite{Prisco00thesis,castro01thesis}. This has not been the case for blockchain-inspired consensus protocols, which are primarily published in non-peer reviewed venues, or just as whitepapers, and often have no formal proof of correctness or security analysis.  Indeed, it was not until 2015---7 years after Bitcoin was first released--- that it was formally proved that Bitcoin \pow is a consensus protocol~\cite{garay2015bitcoin}. 
%
%While this trend is perhaps not surprising given the rapid growth in this area, and the fact that Bitcoin was anonymously published by an academic outsider, it is worrisome that protocols with no formal security guarantees are serving as the backbone for systems that may end up being worth millions of dollars. There have already been many examples of vulnerabilities in deployed cryptocurrencies, and we hope that as this space matures it will shift from a ``buyer's beware'' attitude to more careful considerations of the protocols that are being proposed. 

\subsection{Privacy in consensus}

By their very nature transparent distributed ledgers pose significant privacy challenges with respect to both the information contained in them, as well as the privacy of transactions and their meta-data. The original \bitcoin announcement promised `anonymity' as a property of the new system; however, the weak form of pseudonymity offered can be bypassed by tracing attacks~\cite{DBLP:conf/imc/MeiklejohnPJLMVS13}. %Three key directions have been developed to protect privacy in such systems: permissioned private ledgers; mix-based techniques; and zero-knowledge based techniques. 

Permissioned systems, such as BigchainDB~\cite{mcconaghy2016bigchaindb}, have the ability to protect privacy by restricting the set of core participants in the consensus to a small vetted set---that are assumed to be trusted both for the integrity (safety) of the systems, its liveness, and can also be trusted for keeping secrets (privacy). However, trusting a set of entities for privacy is of a different nature than trusting them for integrity: if information is replicated any node may violate the property, and such a violation cannot be detected within the system---since it only involves leaking secrets. Furthermore, relying on permissioned ledgers for privacy forces the system to rely on closed groups, for reasons besides efficiency of consensus---making this design choice hard to change.

%Since it does not intersect with consensus efficiency, we do not address here techniques related to mixing. 
Protocol-layer techniques for protecting privacy also have implications for scalability. The most established exemplar of this family is Zcash~\cite{DBLP:conf/sp/Ben-SassonCG0MTV14}, where a transaction contains a succinct non-interactive zero-knowledge proof (zk-SNARK)~\cite{DBLP:conf/asiacrypt/Groth10}, proving that it is spending an existing unspent coin, but without publicly specifying which one. The cost of constructing such SNARKs and verifying them, as well as their size,  affects the efficiency and scalability of protocols. 
%For this reason there is a preference for using SNARKs, over traditional zero-knowledge proofs, since they are of constant size, no matter the complexity of the statement proven. However, they are expensive to build, in relation to the argument to prove, raising concerns about efficiency as the number of transactions grow.
Furthermore, since coins are spent in private, it is impossible to prune past transactions to maintain a smaller Unspent Transaction Output (\emph{utxo}), and checkpointing is ineffective. Thus the state necessary to validate transactions grows indefinitely. 

From a systems perspective, such systems expose the minimum amount of information for validation, and to agree on a consensus on the ordering of transactions---while the actual construction and execution of transactions happens off-chain between the parties having visibility into the full secrets. The Chainspace platform applies this privacy pattern to general smart contracts~\cite{chainspace}. Others~\cite{Vukolic:2017:RPB} argue that distributed ledgers can decouple the ordering---performed in public on cryptographic commitments of transactions---from the validation containing private information, that is only checked by a trusted cabal. Such architectures can scale at the same rate as the core ordering protocol, but do not provide any universal end-to-end verifiability.

\section{Conclusions}

\label{conclusion}

%While originally developed in the context of \bitcoin, blockckchains have now independently evolved to cater to a diverse range of applications, thanks to the key properties of resilience, integrity, and transparency.  
%The blockchain enables these properties in a decentralized setting, hence the issue of consensus---that is, how to reach agreement on a value among a group of nodes---lies at its core.
The last few years have seen a dramatic surge in blockchain consensus protocols, 
%both in academia and industry, 
 as a result of which the field has grown increasingly complex. 
We presented a comprehensive systematization of blockchain consensus protocols, and evaluated their performance and security properties using a novel framework.
In a broader context, this work has highlighted a number of open areas and challenges related to: \first gaps between BFT and blockchains, \second security \vs performance tradeoffs, \third incentives, and \fourth privacy.
This longitudinal perspective makes a timely contribution to the prolific and vibrant area of blockchain consensus protocols: the wide-scale adoption of blockchains is constrained by their performance and scalability limitations, and is desperately in need of new and faster consensus protocols that can cater to varying security and privacy requirements. 

\ifproceedings{

\noindent {\bf Acknowledgements.} George Danezis, Shehar Bano and Alberto Sonnino are supported in part by EPSRC Grant EP/M013286/1 and the EU H2020 DECODE project under grant agreement number 732546. Mustafa Al-Bassam is supported by a scholarship from The Alan Turing Institute. Sarah Meiklejohn, Shehar Bano (in part) and Patrick McCorry are supported by EPSRC grant EP/N028104/1.
}\fi

{\footnotesize
\bibliographystyle{abbrv}
\bibliography{ms}
}

\appendix
\subsection{List of Papers}
\label{appendix}

%\begin{table}[t]
\renewcommand{\arraystretch}{1.5}
{\footnotesize
\begin{tabular}{p{1.3cm} | p{6.5cm}}
\toprule
    % ============= pow ============= %
    \textsc{\pow} & 
    \cite{bitcoin}, 
    \cite{hashcash}, 
    \cite{timestampattack1}, \cite{timestampattack2},
    \cite{selfishmining}, 
    \cite{conf/p2p/DeckerW13}, 
    \cite{bitcoinng}, 
    \cite{garay2015bitcoin},
    \cite{dwork1993crypto}, 
    \cite{luu2017smart},
    \cite{gervais2016security},
    \cite{poon2015bitcoin},
    \cite{wood2014ethereum},
    \cite{sompolinsky2013accelerating},
    \cite{luu2017smart},
    \cite{miller2015nonoutsourceable},
    \cite{lernerdecor+},
    \cite{pass2017fruitchains}\\
    
    % ============= pox ============= %
    \textsc{\pox} & 
    \cite{BentovGM14},
    \cite{cryptoeprint:2014:452},
    \cite{ouroboros}, 
    \cite{ouroborospraos},
    \cite{snow-white},
    \cite{peercoin},
    \cite{blackcoin}
    \cite{DBLP:journals/corr/Chepurnoy16}
    \cite{casper}
    \cite{permacoin}
    \cite{Fuchsbauer2015SpaceMintAC}
    \cite{sawtooth}
    \cite{rem}\\
    
    % ============= hybrid ============= %
    \textsc{hybrid} & 
    \cite{mazieres2015stellar},
    \cite{byzcoin},
    \cite{cosi},
    \cite{rscoin},
    \cite{algorand},
    \cite{elastico},
    \cite{chainspace},
    \cite{omniledger},
    \cite{elastico},
    \cite{Vukolic:2017:RPB},
    \cite{solidus},
    \cite{buchman2016tendermint},
    \cite{churyumovbyteball},
    \cite{miller2016honey},
    \cite{pass2017hybrid},
    \cite{abraham2016bvp},
    \cite{barger2017scalable},
    \cite{swirlds},
    \cite{nikitin2017chainiac},
    \cite{swanson2015consensus},
    \cite{hearn2016corda},
    \cite{guocrysto},
    \cite{laurie2011efficient},
    \cite{ren2017implicit},
    \cite{cachin2016non},
    \cite{Decker:2016},
    %\cite{liu2016scalable}, %\bano{good reference for the relevant discussion point.} 
    %\cite{cachin2001distributing},
    %\cite{decker2016bitcoin},
    \cite{wood2016polkadot},
    \cite{ren2017practical},
    \cite{schwartz2014ripple},
    %\cite{jia2017robust},
    \cite{gencer2016service},
    \cite{martino2016kadena},
    \cite{li2017towards},
    \cite{cachin2016architecture},
    \cite{mcconaghy2016bigchaindb} \\

\bottomrule
\end{tabular}
%\caption{List of all paper read for this SoK.}
%\label{tab:list_of_papers}
%\end{table}
}

\subsection{Glossary}

\label{glossary}

\begin{itemize}

\item \emph{Adversary model}: The fraction of malicious or faulty nodes that the consensus protocol can tolerate (\ie it will operate correctly despite the presence of such nodes).

\item \emph{Code available}: Whether the code implementing the system is publicly available.

\item \emph{Committee}: How the participants work together to participate in the consensus protocol; either they all work together (single committee), or they are divided in multiple subgroups (multiple committees).

\item \emph{Committee Formation}: How the members of the committee are chosen, for example \via proof-of-work, proof-of-stake, trusted hardware~\etc

\item \emph{Consistency}: The likelihood that the system will reach consensus on a proposed value; it can be either strong or weak.

\item \emph{\dos resistance}: Resilience of the node(s) involved in consensus to denial-of-service (\dos) attacks. If the participants of the consensus protocol are known in advance, an adversary may launch a \dos attack against them.

\item \emph{Experimental setup}: The configuration used to generate the numbers reported for throughput and latency.

\item \emph{Incentives}: The mechanisms that keep nodes motivated to participate in the system and follow its rules.

\item \emph{Inter-committee Configuration}: How the members are assigned to the committee in a single committee setting; either members serve on the committee permanently (static), or they are changed at regular intervals (rolling, or full swap).

\item \emph{Inter-committee Consensus}: Reaching agreement on a value among nodes in a single committee.

\item \emph{Intra-committee configuration}: How the members are assigned to the committees in a multiple committees setting; it can be static or dynamic.

\item \emph{Intra-committee Consensus}: Reaching agreement on a value among nodes across multiple committees; this can be optionally mediated by an external party (\eg the client).

\item \emph{Latency}: The time it takes from when a transaction is proposed until consensus has been reached on it.

%\item \emph{Correctness proof availability} captures if a formal proof of correctness has been provided for the consensus protocol. 

\item \emph{Leader}: The leader of the consensus protocol, which can be either elected among the current committee (internally), externally, or flexibly (\eg through arbitrary smart contracts).

\item \emph{Participants}: The nodes that participate in the consensus protocol.

\item \emph{Permissioned blockchain}: Only participants selected by the appropriate authorities can participate in the consensus protocol.

\item \emph{Permissionless blockchain}: Anyone can join the system and participate in the consensus protocol.

\item \emph{Scalability}: The system's ability to achieve greater throughput when consensus involves a larger number of nodes.

\item \emph{Throughput}: The maximum rate at which transactions can be agreed upon by the consensus protocol (transactions per second/hour).

\item \emph{Transaction censorship resistance}: The system's resilience to proposed transactions being suppressed (\ie censored) by malicious node(s) involved in consensus.

\end{itemize}

\end{document}